%% file: gosam_main.tex
\documentclass[twocolumn,epjc3]{svjour3}
\usepackage{amsmath,amssymb,amsfonts}
\usepackage{graphicx}
\usepackage{color}
\usepackage{subfigure}
\usepackage{units}
\usepackage{listings}

\definecolor{shaded}{RGB}{210,210,210}
\newcommand{\GOLEM}{{\textsc{Go\-Sam}}}
\newcommand{\GOSAM}{{\textsc{Go\-Sam}}}

\newcommand{\SHERPA}{{\textsc{SHER\-PA}}}
\newcommand{\MCFM}{{\textsc{MCFM}}}
\newcommand{\POWHEG}{{\texttt{POW\-HEG}}}
\newcommand{\QGRAF}{{\texttt{QGRAF}}}
\newcommand{\FORM}{{\texttt{FORM}}}
\newcommand{\SPINNEY}{{\texttt{spin\-ney}}}
\newcommand{\HAGGIES}{{\texttt{hag\-gies}}}
\newcommand{\SAMURAI}{{\textsc{Sa\-mu\-rai}}}
\newcommand{\GOLEMVC}{{\texttt{Go\-lem95C}}}
\newcommand{\PJFRY}{{\texttt{PJFRY}}}
\newcommand{\FORTRAN}{{\texttt{For\-tran}}}
\newcommand{\PYTHON}{{\texttt{Py\-thon}}}
\newcommand{\FEYNRULES}{{\texttt{Feyn\-Rules}}}
\newcommand{\UFO}{{\texttt{UFO}}}
\newcommand{\LANHEP}{{Lan\-HEP}}
\newcommand{\LOOPTOOLS}{{Loop\-Tools}}
\newcommand{\CUTTOOLS}{{Cut\-Tools}}
\newcommand{\tHV}{{'t\,Hooft Veltman}}

\newcommand{\fmslash}[1]{{\ensuremath{/\!\!\!{#1}}}}
\newenvironment{kinematics}{%
\fontsize{8pt}{10pt}\selectfont%
\begin{tabular}{l|rrrr}
   & \multicolumn{1}{c}{$E$}
   & \multicolumn{1}{c}{$p_x$} 
   & \multicolumn{1}{c}{$p_y$}
   & \multicolumn{1}{c}{$p_z$} \\
\hline}{%
\end{tabular}}

\newenvironment{parameters}{%
\begin{tabular}{l@{\hspace{1.5em}}r|l@{\hspace{1.5em}}r}
\hline
\multicolumn{4}{c}{parameters}\\
\hline
}{%
\end{tabular}}

\newcommand{\bea}{\begin{eqnarray*}}
\newcommand{\eea}{\end{eqnarray*}\noindent}
\newcommand{\bcen}{\begin{center}}
\newcommand{\ecen}{\end{center}}

\journalname{Eur. Phys. J. C}

\begin{document}

\title{Automated One-Loop Calculations with \GOSAM{}}
\author{%
       Gavin~Cullen\thanksref{cullen,desy,edi}
  \and Nicolas~Greiner\thanksref{greiner,mpp,uiuc}
  \and Gudrun~Heinrich\thanksref{heinrich,mpp}
  \and Gionata~Luisoni\thanksref{luisoni,ippp}
  \and Pierpaolo~Mastrolia\thanksref{mastrolia,mpp,padova}
  \and Giovanni~Ossola\thanksref{ossola,ny,ny2}
  \and Thomas~Reiter\thanksref{reiter,mpp}
  \and Francesco~Tramontano\thanksref{tramontano,cern}
}


\institute{%
     DESY, Zeuthen, Germany\label{desy}
\and School of Physics and Astronomy, The University of Edinburgh,
     UK\label{edi}
\and Max-Planck-Institut f\"ur Physik, M\"unchen, Germany\label{mpp}
\and Department of Physics, University of Illinois
     at Urbana-Champaign\label{uiuc}
\and Institute for Particle Physics Phenomenology, University of Durham,
     UK\label{ippp}
\and Dipartimento di Fisica, Universit\`a di Padova, Italy\label{padova}
\and New York City College of Technology, City University of New York\label{ny}
\and The Graduate School and University Center, City University of New York%
     \label{ny2}
\and CERN, Geneva, Switzerland\label{cern}
}

\date{Received: date / Accepted: date}

\maketitle

\begin{abstract}
We present the program package \GOSAM{} which is designed for the automated calculation of 
one-loop amplitudes for multi-particle processes
in renorma\-lisable quantum field theories. 
The amplitudes, which are generated in terms of Feynman diagrams,
can be reduced using either D-dimensional integrand-level decomposition
or tensor reduction. 
\GOSAM{} can be used to calculate one-loop QCD and/or electroweak corrections to Standard Model processes
and offers the flexibility to link model files for theories Beyond the Standard Model.
A standard interface to programs calculating real radiation is
also implemented. We demonstrate the flexibility of the program
by presenting examples of processes with up to six external legs
attached to the~loop.
\end{abstract}

\newpage
\tableofcontents
\newpage

\section{Introduction}
\label{sec:intro}
\input intro

\section{Overview and Algorithms}
\label{sec:overalg}
\subsection{Overview}
\label{ssec:overview}
\input overview

\subsection{Generation and Organisation of the Diagrams}
\label{ssec:diagen}
\input diagen

\subsection{Algebraic Processing}
\label{ssec:algebra}
\input algebra

\subsection{Code Generation}
\label{ssec:codegen}
\input codegen

\subsection{Conventions of the Amplitudes}
\label{ssec:conventions}
\input conventions

\section{Requirements and Installation}
\label{sec:install}
\input install

\section{Using \GOSAM}
\label{sec:usage}

\subsection{Setting up a simple Process}
\label{ssec:usage}
\input usage

\subsection{Interfacing the code}
\label{ssec:api}
\input api

\subsection{Using the BLHA Interface}
\label{ssec:blha}
\input blha

\subsection{Using External Model Files}
\label{ssec:model}
\input model


\section{Sample Calculations and Benchmarks}
\label{ssec:processes}
The codes produced by \GOSAM{} have been tested on several processes.
In this section we describe some examples of applications. Additional results, whose corresponding code is 
also included in the official distribution of the program, will be reported in Appendix~B. 

\subsection{\texorpdfstring{$pp\to W^{-}+j$}{p p -{}-> W j} with \SHERPA}
\input wjet

\subsection{\texorpdfstring{$pp\to W^\pm+j$}{p p -{}-> W j}, EW Corrections}
\input wjetew

\subsection{\texorpdfstring{$\gamma\gamma\to\gamma\gamma$}{%
photon photon -{}-> photon photon}}
\input yyyy

\subsection{\texorpdfstring{$pp\to\chi^0_1\chi^0_1$}{%
p p -{}-> chi\^{ }0\_1 chi\^{ }0\_1} in the MSSM}
\input neutralino

\subsection{\texorpdfstring{$e^+e^-\to e^+e^-\gamma$}{%
e+ e- -{}-> e+ e- photon} in QED}
\input eeey

\subsection{\texorpdfstring{$pp\to t\overline{t}H$}{%
pp -{}-> t t-bar H}}
\input ttH

\subsection{\texorpdfstring{$gg\to t\overline{t}Z$}{%
pp -{}-> t t-bar Z}}
\input ttZ

\subsection{\texorpdfstring{$pp\to b\overline{b}b\overline{b}+X$}{%
p p -{}-> b b-bar b b-bar + X}}
\input bbbb
\subsection{\texorpdfstring{$pp\to t\overline{t}b\overline{b}+X$}{%
p p -{}-> t t-bar b b-bar + X}}
\input ttbb

\subsection{\texorpdfstring{$pp\to W^+W^- b\overline{b}$}{%
pp -{}-> W+ W- b b-bar}}
\label{sec:wwbb}
\input wwbb

\subsection{\texorpdfstring{$u\overline{d}\to W^+ggg$}{%
u d-bar -{}-> W+ g g g}}
\input udwggg

\subsection{\texorpdfstring{$u {\bar d} \to W^+(\to \nu_e e^+) b\overline{b}$ (massive $b$-quark)}{%
u d-bar -{}-> W+ b b-bar}}
\input wBB

\section{Conclusions}
\label{sec:conclusion}
\input conclusion

\begin{acknowledgements}
We would like to thank the \SHERPA{} collaboration for the support, in particular Jennifer Archibald, Frank Krauss and Marek Sch\"onherr.
We also would like to thank Rikkert Frederix, Adam Kardos, Stefano Pozzorini, Zoltan Trocsanyi, 
Thomas Schutzmeier and Christian Sturm for their input to various comparisons and clarifying discussions, 
and Edoardo Mirabella for important feedback on Samurai.
G.C. and G.L. were supported by the British Science and Technology Facilities
Council (STFC).
The work of G.C was supported by DFG
Sonderforschungsbereich Transregio 9, Computergest\"utzte Theoretische Teilchenphysik.
N.G. was supported in part by the U.S. Department of Energy under contract
No. DE-FG02-91ER40677.
P.M. and T.R. were supported by the Alexander von
Humboldt Foundation, in the framework of the Sofja Kovaleskaja Award Project
``Advanced Mathematical Methods for Particle Physics'', endowed by the German
Federal Ministry of Education and Research.
The work of G.O. was supported in part by the National Science Foundation
under Grant PHY-0855489 and PHY-1068550.
The research of F.T. is supported by Marie-Curie-IEF, project:
``SAMURAI-Apps''.
We also acknowledge the support of the Research Executive Agency (REA)
of the European Union under the Grant Agreement number
PITN-GA-2010-264564 (LHCPhenoNet).
\end{acknowledgements}

\appendix
\renewcommand \thesection{\Alph{section}}
\renewcommand{\theequation}{\Alph{section}.\arabic{equation}}
\setcounter{equation}{0}

\section*{Appendix}

\section{Examples included in the release}
\label{ssec:examples}
\input examples

\section{Explicit reduction of $R_2$ rational terms}
\label{ssec:app-r2}
\input app-r2

\bibliographystyle{spphys}


\end{document}

%% file: intro.tex
The Standard Model is currently being re-discovered at the LHC, and 
new exclusion limits on Beyond the Standard Model particles -- 
and on the Higgs mass -- are being delivered by the experimental 
collaborations with an impressive speed. 
Higher order corrections play an important role in 
obtaining bounds on the Higgs boson and New Physics. In particular, the exclusion limits for the Higgs boson would look 
very different if we only had leading order tools at hand. 
Further, it will be very important to have precise theory predictions  
to 
constrain model parameters once a signal of New Physics has been established. 
Therefore it is of major importance to provide tools 
for next-to-leading order (NLO) predictions which are largely automated, 
such that signal and background rates for a multitude of processes 
can be estimated reliably. 

The need for an automation of NLO calculations has been noticed some time ago
and lead to public programs like FeynArts\,\cite{Hahn:2000kx} and
 QGraf\,\cite{Nogueira:1991ex} 
for diagram generation and FormCalc/LoopTools\,\cite{Hahn:1998yk} and 
{\small GRACE} \,\cite{Belanger:2003sd} for 
the automated calculation of NLO corrections, primarily in the electroweak sector. 
However, the calculation of one-loop amplitudes with more than four external legs  
were still tedious case-by-case calculations. Only very recently, conceptual 
and technical advances in multi-leg one-loop calculations allowed the  calculation 
of six-point\,\cite{Denner:2005fg,Berger:2009ep,Berger:2009zg,KeithEllis:2009bu,Melnikov:2009wh,Berger:2010vm,Campbell:2010cz,Bredenstein:2009aj,Bredenstein:2010rs,Bevilacqua:2009zn,Bevilacqua:2010ve,Binoth:2009rv,Greiner:2011mp,Bevilacqua:2010qb,Denner:2010jp,Melia:2010bm,Melnikov:2010iu,Campanario:2011ud,Frederix:2010ne,Cascioli:2011va} 
  and even seven-point\,\cite{Berger:2010zx,Ita:2011wn} processes 
at all, and opened the door to the possibility of an {\it automated} generation 
and evaluation of multi-leg one-loop amplitudes.
As a consequence, already existing excellent public tools, each containing
a collection of hard-coded individual processes, like  e.g. 
MCFM\,\cite{Campbell:1999ah,Campbell:2011bn}, VBFNLO\,\cite{Arnold:2008rz,Arnold:2011wj}, 
MC@NLO\,\cite{Frixione:2002ik,Frixione:2010wd}, POWHEG-\hspace{0pt}Box \cite{Frixione:2007vw,Alioli:2010xd}, POWHEL\,\cite{Kardos:2011qa,Garzelli:2011vp,Kardos:2011na},
can be flanked by flexible automated tools 
such that basically any process which may turn out to be important for 
the comparison of LHC findings to theory 
can be evaluated at NLO accuracy.

We have recently experienced major advances in the activity of constructing packages 
for fully automated one-loop calculations, see e.g.
\cite{vanHameren:2009dr,Hirschi:2011pa,Mastrolia:2010nb,%
Cullen:2010hz,Bevilacqua:2011xh,Reina:2011mb}.
The concepts that lead to these advances have been recently reviewed in \cite{Ellis:2011cr}.
Among the most important developments are   
the integrand-reduction technique\,\cite{Ossola:2006us,Ossola:2007bb} and the 
generalized $n$-dimensional unitarity\,\cite{Ellis:2008ir}. 
Their main outcome is a numerical reconstruction of a representation of 
the tensor structure of any one-loop integrand where the multi-particle pole configuration is manifest. 
As a consequence, decomposing one-loop amplitudes in terms of basic integrals becomes 
equivalent to reconstructing the polynomial forms of the residues to all multi-particle cuts.
Within this algorithm, the integrand of a given scattering amplitude, 
carrying complete and explicit information on the chosen dimensional-regularisation scheme, 
is the only input required to accomplish the task of its evaluation. In fact, 
the integration is substituted by a much simpler operation, namely by polynomial fitting, 
which requires the sampling of the integrand on the solutions of generalised on-shell conditions.

In this article, we present the program package \GOLEM{} which allows the automated calculation of 
one-loop amplitudes for multi-particle processes. 
Amplitudes are expressed in terms of Feynman diagrams, where the integrand is generated analytically using \QGRAF~\cite{Nogueira:1991ex}, \FORM~\cite{Vermaseren:2000nd},
\SPINNEY~\cite{Cullen:2010jv} and \HAGGIES~\cite{Reiter:2009ts}. 
The individual program tasks are steered via python scripts, while the user only needs to edit an ``input card" to specify the details of the process to be calculated,
and launch the generation of the source code and its compilation, without having to worry about internal details of the 
code generation.

The program offers the option to use different reduction techniques:
either the unitarity-based integrand reduction 
as implemented in \SAMURAI~\cite{Mastrolia:2010nb} 
or traditional tensor reduction as implemented in
\GOLEMVC~\cite{Binoth:2008uq,Cullen:2011kv} interfaced through
tensorial reconstruction at the integrand level~\cite{Heinrich:2010ax},
or a combination of both.
It can be used to calculate one-loop corrections within both QCD and electroweak theory. 
Beyond the Standard Model theories can be interfaced using
FeynRules~\cite{Degrande:2011ua} or \LANHEP~\cite{Semenov:2010qt}.
The Binoth-Les Houches-interface\,\cite{Binoth:2010xt} to programs providing the real radiation 
contributions is also included.

The advantage of generating analytic expressions for the integrand of each diagram gives the user the flexibility 
to organize the computation according to his own efficiency preferences.
For instance, the computing algorithm can proceed either diagram-by-diagram or by grouping diagrams that share  
a common set of denominators (suitable for a unitarity-based reduction), and it can deal with 
the evaluation of the rational terms either on the same footing as the rest of the amplitude, or through an independent 
routine which evaluates them analytically. 
These options and the other features of \GOLEM{} will be discussed in detail in the following.

In Section \ref{sec:overalg}, after giving an overview on the diagram generation and 
on processing gauge-group and Lorentz algebra,
we discuss the code generation and the reduction strategies. 
The installation requirements are given in Section \ref{sec:install}, 
while Section \ref{sec:usage} describes the usage of \GOLEM{}, containing all the 
set-up options which can be activated 
by editing the input card. In Section \ref{ssec:processes} we show results for processes of various complexity. 
The release of \GOLEM{} is accompanied by the generated code for some example processes, listed 
in Appendix \ref{ssec:examples}.

%% file: overview.tex
\GOSAM{} produces, in a fully automated way, all the code required to perform
the calculation of one-loop matrix elements. 
There are three main steps in the process of constructing the code:
the generation of all contributing diagrams within a process directory, 
the generation of the \FORTRAN{} code, 
and finally compiling and linking the generated code.
These steps are self-contained in the sense that after each
step all the files contained in the process directory could be transfered to a different machine
where the next step will be carried out.

In the following sections we focus on the algorithms that are employed for the
construction of the code to produce and evaluate matrix elements.

The first step (\emph{setting up a process directory}), which consists in the
generation of some general source files and the generation of the diagrams,
is described in Section~\ref{ssec:diagen}. The second step 
(\emph{generating the fortran code}) is carried out by means of advanced algorithms
for algebraic manipulation and code optimization which are presented in Sections \ref{ssec:algebra}
and~\ref{ssec:codegen}. The third step (\emph{compilation and linking}) is not 
specific to our code generation, 
therefore will not be described here.

The practical procedures to be followed by the user in generating the code will 
be given in Section~\ref{sec:usage}, which can be considered a short version of the user manual.

%% file: diagen.tex
\begin{figure}
\centering
\subfigure[Diagram~1.]{\label{fig:qgrafex01:scaleless}%
\includegraphics[width=0.45\columnwidth]{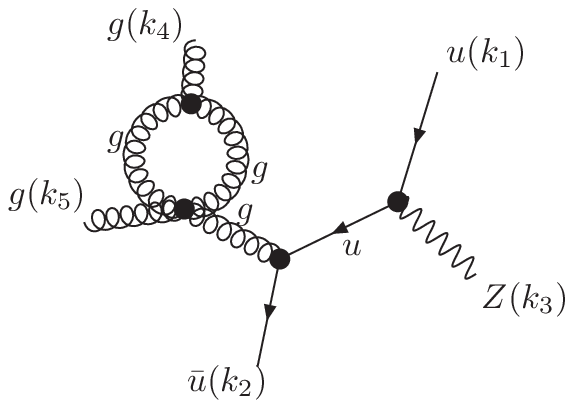}}
\subfigure[Diagram~2.]{\label{fig:qgrafex01:wbox}%
\includegraphics[width=0.45\columnwidth]{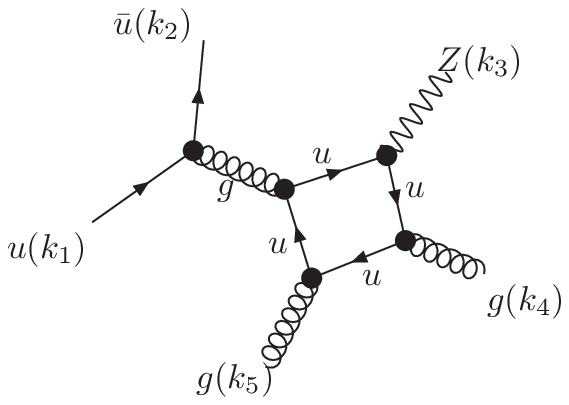}}
\caption{Two examples for diagrams which are difficult to isolate using
\QGRAF{}. The diagram in Fig.~\ref{fig:qgrafex01:scaleless} is zero in dimensional
regularisation. However, in \QGRAF{} there is no operator to identify this
type of diagrams.
In Fig.~\ref{fig:qgrafex01:wbox} the $Z$ boson is emitted from a closed
quark line. These diagrams form a separate gauge invariant class and could
be treated separately from diagrams where the $Z$ boson comes from an
external quark line.}
\label{fig:qgrafex01}
\end{figure}

For the diagram generation both at tree level and one-loop level
we employ the program \QGRAF~\cite{Nogueira:1991ex}. This program
already offers several ways of excluding unwanted diagrams, for example
by requesting a certain number of propagators or vertices of a certain type
or by specifying topological properties such as the presence of tadpoles or
on-shell propagators.
Although \QGRAF{} is a very reliable and fast generator, we extend its
possibilities by adding another level of analysing and filtering over diagrams
by means of \PYTHON{}. 
This gives several advantages: first of all, the possibilities offered by \QGRAF{} are not always sufficient
to distinguish certain classes of diagrams (see examples in Fig.~\ref{fig:qgrafex01});
secondly, \QGRAF{} cannot handle the sign for diagrams with Majorana fermions
in a reliable way; finally, in order to fully optimize 
the reduction, we want to classify and group diagrams 
according to the sets of their propagators.

Within our framework, \QGRAF{} generates three sets
of output files: an expression for each diagram to be processed with
\FORM~\cite{Vermaseren:2000nd}, \PYTHON{} code for drawing all diagrams,
and \PYTHON{} code for computing the properties of each diagram. 
The information about the model
for \QGRAF{} is either read from the built-in Standard Model file or
is generated from a user defined \LANHEP~\cite{Semenov:2010qt}
or Universal FeynRules Output (\UFO)~\cite{Degrande:2011ua} file.

The \PYTHON{} program automatically performs several operations:
\begin{itemize}
\item diagrams whose color factor turns out to be zero are dropped
      automatically;
\item the fermion flow is determined and used to compute an overall
      sign for each diagram, which is relevant in the presence of
      Majorana fermions;
\item the number of propagators containing the loop momentum, i.e.
      the loop size of the diagram, the tensor rank and the kinematic
      invariants of the associated loop integral are computed;
\item diagrams with an associated vanishing loop integral
      (see Fig.~\ref{fig:qgrafex01:scaleless}) are detected
      and flagged for the diagram selection;
\item all propagators and vertices are classified for the diagram selection;
      diagrams containing massive quark self-energy insertions or
      closed massless quark loops are specially flagged.
\end{itemize}
Any one-loop diagram can be written in the form  
\begin{equation}
\mathcal{D}=\int\!\!\frac{\mathrm{d}^n q}{i\pi^{n/2}}%
\frac{\mathcal{N}(q)}{\prod_{l=1}^N\left[(q+r_l)^2-m_l^2+i\delta\right]}
\end{equation}
where the numerator is a polynomial of tensor\footnote{
Index contractions in Eq.~\eqref{eq:diagen:numer} are
understood in $n$-dimensional space.} rank~$r$.
\begin{align}
\label{eq:diagen:numer}
\mathcal{N}(q)&=C_0+C_1^{\mu_1}q_{\mu_1}+\ldots+
  C_r^{\mu_1\ldots\mu_r}q_{\mu_1}\cdots q_{\mu_r}\, , \\
\intertext{and the $N\times N$ kinematic matrix is defined as}
  S_{ij}&=(r_i-r_j)^2 - m_i^2 - m_j^2\, .
\end{align}
All masses can be either real or complex.
Important information about the integrals that will appear in the reduction
of each one-loop diagram is contained in the tensor rank~$r$ of the loop 
integral and its kinematic matrix~$S_{ij}$.

We define a preorder relation
on one-loop diagrams, such that $\mathcal{D}_1\preceq\mathcal{D}_2$
if their associated matrices $S(\mathcal{D}_1)$ and $S(\mathcal{D}_2)$
are related by a finite (not necessarily unique) chain of transformations
\begin{equation}
S(\mathcal{D}_2)\stackrel{T_1}{\longrightarrow}S^\prime
\stackrel{T_2}{\longrightarrow}\ldots\stackrel{T_m}{\longrightarrow}
S(\mathcal{D}_1)
\end{equation}
where each transformation is one of the following:
\begin{itemize}
\item the identity,
\item the simultaneous permutation of rows and columns,
\item the simultaneous deletion of the row and column with the same
      index, which corresponds to \emph{pinching} the corresponding
      propagator in the diagram.
\end{itemize}
The relation ``$\preceq$'' can be read as ``appears in the reduction of''.
Our algorithm groups the one-loop diagrams
$\mathcal{D}_1,\ldots, \mathcal{D}_D$ of a process
into subsets $V_1,\ldots, V_G$ such that
\begin{itemize}
\item $V_1,\ldots, V_G$ form a partition of 
$\{\mathcal{D}_1, \ldots \mathcal{D}_D\}$ and
\item each cell $V_i$ contains a maximum element $\max V_i\in V_i$,
      such that $\mathcal{D}\preceq\max V_i\, , \forall\,\mathcal{D}\in V_i$.
\end{itemize}
The partitioning procedure provides an important gain in efficiency, because while
carrying out the tensor reduction for the diagram $\max V_i$, all other
diagrams in the same cell $V_i$ are reduced with virtually no additional
computational cost. The gain in efficiency can be observed when reducing 
the diagram using the OPP method~\cite{Ossola:2006us}
and its implementations in \CUTTOOLS~\cite{Ossola:2007ax} and
\SAMURAI~\cite{Mastrolia:2010nb}, as well as
in classical tensor reduction methods as implemented e.g. 
in \GOLEMVC~\cite{Binoth:2008uq,Cullen:2011kv}, \PJFRY~\cite{Fleischer:2010sq}
and \LOOPTOOLS~\cite{Hahn:1998yk,vanOldenborgh:1989wn}.

In order to draw the diagrams, we first compute an ordering of the external legs
which allows for a planar embedding of the graph. Such ordering can always be found
for a tree or a one-loop graph since non-planar graphs only start to appear in diagrams
with two or more loops.
After the legs have been assigned to the vertices 
of a regular polygon, we use our own implementation of the algorithms described in~\cite{Ohl:1995kr} for fixing
the coordinates of the remaining vertices; the algorithm has been extended to determine an
appealing layout also for graphs containing tadpoles. Starting from these coordinates and 
using the package Axodraw~\cite{Vermaseren:1994je},
\GOSAM{} generates a \LaTeX{} file that contains graphical
representations of all diagrams.

%% file: algebra.tex
\subsubsection{Color Algebra}
In the models used by \GOSAM{}, we allow one unbroken gauge group
$\mathsf{SU}(N_C)$ to be treated implicitly; any additional gauge group,
broken or unbroken, needs to be expanded explicitly. Any particle
of the model may be charged under the $\mathsf{SU}(N_C)$ group in the
trivial, (anti-)fundamental or adjoint representation. Other representations
are currently not implemented.

For a given process we project each Feynman diagram onto a color basis
consisting of strings of generators
   $T^{A_1}_{ii_1}T^{A_2}_{i_1i_2}\cdots T^{A_p}_{i_{p-1}j}$ and
Kronecker deltas $\delta_{ij}$ but no contractions of adjoint indices
and no structure constants $f^{ABC}$. Considering, for example, the
process
\begin{displaymath}
u(1)+\bar{u}(2)\rightarrow Z(3)+g(4)+g(5)
\end{displaymath}
\GOLEM{} finds the color basis
\begin{align*}
\vert c_{1}\rangle &= q^{(1)}_{i_1}\bar{q}^{(2)}_{j_2}g_{(4)}^{A_4}g_{(5)}^{A_5}(T^{A_4}T^{A_5})_{j_2i_1},\\
\vert c_{2}\rangle &= q^{(1)}_{i_1}\bar{q}^{(2)}_{j_2}g_{(4)}^{A_4}g_{(5)}^{A_5}(T^{A_5}T^{A_4})_{j_2i_1},\\
\vert c_{3}\rangle &= q^{(1)}_{i_1}\bar{q}^{(2)}_{j_2}g_{(4)}^{A_4}g_{(5)}^{A_5}\delta_{j_2i_1}\mathrm{tr}\{T^{A_5}T^{A_4}\},
\end{align*}
where $q_{i_\bullet}^{(\bullet)}$ and $g_{(\bullet)}^{A_\bullet}$ are
the color parts of the quark and gluon wave functions respectively.
The dimension of this color basis for $N_g$ external gluons and
$N_{q\bar{q}}$ quark-antiquark pairs is given by~\cite{Reiter:2009kb}:
\begin{equation}
d(N_g,N_{q\bar{q}}) = \sum_{i=0}^{N_g} (-1)^i\left(\begin{array}{c}N_g\\i%
\end{array}\right)
\cdot\left(N_g+N_{q\bar{q}}-i\right)!\;.
\end{equation}
It should be noted that the color basis constructed in this way is not
a basis in the mathematical sense, as one can find linear relations between
the vectors $\vert c_i\rangle$ once the number of external partons is large
enough.

Any Feynman diagram can be reduced to the form
\begin{equation}
\mathcal{D}=\sum_{i=1}^k \mathcal{C}_i \vert c_{i}\rangle
\end{equation}
for the process specific color basis $\vert c_{1}\rangle,\ldots,%
\vert c_{k}\rangle$ by applying the following set of relations:
\begin{align}
T^A_{ij}T^A_{kl} &=T_R\left(\delta_{il}\delta_{kj}
                    - \frac1{N_C}\delta_{ij}\delta_{kl}\right),\\
f^{ABC} &=\frac{1}{iT_R}\left(%
   T^A_{ij}T^B_{jk}T^C_{ki}-T^A_{ij}T^C_{jk}T^B_{ki}\right).
\end{align}
The same set of simplifications is used to compute the matrices
$\langle c_i\vert c_j\rangle$ and
$\langle c_i\vert T_I\cdot T_J\vert c_j\rangle$. The former is needed
for squaring the matrix element, whereas the latter is used to provide
color correlated Born matrix elements which we use for checking the
IR poles of the virtual amplitude and also to provide the relevant
information for parton showers like
\POWHEG~\cite{Nason:2004rx,Frixione:2007vw,Alioli:2010xd}.
For the above example, \GOLEM{} obtains\footnote{%
In the actual code the results are given in terms of $T_R$
and~$N_C$ only.}
\begin{equation}
\langle c_i\vert c_j\rangle=T_R C_F\left(\begin{array}{ccc}%
(N_C^2-1) & -1 & N_C \\
-1 & (N_C^2-1) & N_C \\
N_C & N_C & N_C^2
\end{array}\right).
\end{equation}
Similarly, the program computes the matrices $\langle c_i\vert T_I\cdot T_J%
\vert c_j\rangle$ for all pairs of partons $I$ and $J$.

If $\mathcal{M}^{(0)}$ denotes the tree-level matrix element of the
process and we have
\begin{equation}
\mathcal{M}^{(0)}=\sum_{j=1}^k \mathcal{C}^{(0)}_j \vert c_j\rangle,
\end{equation}
then the square of the tree level amplitude can be written as
\begin{equation}
\left\vert\mathcal{M}^{(0)}\right\vert^2=\sum_{i,j=1}^k
\left(\mathcal{C}^{(0)}_i\right)^\ast\mathcal{C}^{(0)}_j
\langle c_i\vert c_j\rangle.
\end{equation}
For the interference term between leading and next-to-leading order
we use a slightly different philosophy. First of all we note that it is
sufficient to focus on a single group $V_\alpha$ as defined in
Section~\ref{ssec:diagen},
\begin{multline}
\left(\mathcal{M}^{(1)}\right)^\dagger\mathcal{M}^{(0)}+h.c.=\\
\sum_\alpha \int\!\!\frac{\mathrm{d}^n q}{i\pi^{n/2}}
\frac{\mathcal{N}_\alpha(q)}{%
\prod_{l=1}^N\left[(q+r_l)^2-m_l^2+i\delta\right]}
+ h.c.
\end{multline}
In order to reduce the complexity at the level of the reduction, we perform
the contraction with the tree-level already at the integrand level,
\begin{equation}
\mathcal{N}_\alpha(q)=\sum_{i,j=1}^k
\langle c_i\vert c_j\rangle
\left(\mathcal{C}^{(0)}_i\right)^\ast\mathcal{C}^{(1)}_j(q),
\end{equation}
where $\mathcal{C}_j^{(1)}$ is formed by the sum over the corresponding
coefficients of all diagrams $\mathcal{D}\in V_\alpha$.

\subsubsection{Lorentz Algebra}
In this Section we discuss the algorithms used by \GOSAM{} 
to transform the coefficients $\mathcal{C}^{(0)}_i$
and $\mathcal{C}^{(1)}_i(q)$, as defined in the previous section,
such that the result is suitable for efficient numerical evaluation.
One of the major goals is to split the $n$-dimensional
algebra ($n=4-2\varepsilon$) into strictly four-dimensional 
objects and symbols representing
the higher-dimensional remainder.

In \GOLEM{} we have implemented the 't~Hooft-\hspace{0pt}Veltman
scheme (HV) and
dimensional reduction~(DRED). In both schemes all external vectors
(momenta and polarisation vectors) are kept in four dimensions.
Internal vectors, however, are kept in the $n$-dimensional vector space.
We adopt the conventions used in~\cite{Cullen:2010jv}, where
$\hat{k}$ denotes the four-dimensional projection of an in general
$n$-dimensional vector $k$. The $(n-4)$-dimensional orthogonal projection
is denoted as~$\tilde{k}$. For the integration momentum $q$ we introduce
in addition the symbol $\mu^2=-\tilde{q}^2$, such that
\begin{equation}
q^2=\hat{q}^2+\tilde{q}^2=\hat{q}^2-\mu^2.
\end{equation}
We also introduce suitable projectors by splitting the metric tensor
\begin{equation}
g^{\mu\nu}=\hat{g}^{\mu\nu}+\tilde{g}^{\mu\nu},\quad%
\hat{g}^{\mu\nu}\tilde{g}_{\nu\rho}=0,\quad%
\hat{g}^\mu_\mu=4,\quad\tilde{g}^\mu_\mu=n-4.
\end{equation}

In the follwing, we describe the 't~Hooft algebra in detail. For DRED, the only
differences are that the numerator algebra is performed
in four dimensions for both external and internal vectors
(i.e. $q\equiv\hat{q}$) and that in
the very end all appearances of $q^2$ are replaced by $\hat{q}^2-\mu^2$.

\paragraph{Wave Functions and Propagators} \GOLEM{} contains a library
of representations of wave functions and propagators up to
spin~two\footnote{%
Processes with particles of spin~\nicefrac{3}{2} and spin~2 have not
been tested extensively. Furthermore, these processes can lead to integrals
where the rank is higher than the loop size, which at the moment are neither
implemented  in \SAMURAI{} nor in \GOLEMVC{}.}.
The exact form of the interaction vertices is taken from the model files.

The representation of all wave functions with non-trivial spin is based
on massless spinors. Each massive external vector $p_i$ is replaced by
its light-cone projection~$l_i$
with respect to a lightlike reference vector~$k$,
\begin{equation}\label{eq:light-cone-splitting}
p_i^\mu = l_i^\mu + \frac{p_i^2}{2 p_i\cdot k}k^\mu.
\end{equation}

\begin{table*}
\centering
\subfigure[Assignment of initial and final states for quarks and leptons.]{%
\label{stab:wf-fermions:a}%
\begin{tabular}{l|cc}
     &     $l^-$, $q$ & $l^+$, $\bar{q}$\\
\hline
initial & $u_\alpha(k, j_3)$ & $\bar{v}_\alpha(k, j_3)$ \\
final & $\bar{u}_\alpha(k, j_3)$ & $v_\alpha(k, j_3)$
\end{tabular}
}\;
\subfigure[Wave functions for massless~fermions.]{\label{stab:wf-fermions:b}%
\begin{tabular}{l}
$u_\alpha(k,+1)=v_\alpha(k,-1)=\left\vert k\right\rangle$ \\
$u_\alpha(k,-1)=v_\alpha(k,+1)=\left\vert k\right]$ \\
$\bar{u}_\alpha(k,+1)=\bar{v}_\alpha(k,-1)=\left[k\right\vert$ \\
$\bar{u}_\alpha(k,-1)=\bar{v}_\alpha(k,+1)=\left\langle k\right\vert$
\end{tabular}
}\;
\subfigure[Wave functions for massive~fermions.]{\label{stab:wf-fermions:c}%
\begin{tabular}{ll}
$u_\alpha(p,+1)=\left\vert p^+\right\rangle$ &
$\bar{u}_\alpha(p,+1)=\left[ p^+\right\vert$ \\
$u_\alpha(p,-1)=\left\vert p^+\right]$ &
$\bar{u}_\alpha(p,-1)=\left\langle p^+\right\vert$ \\
$v_\alpha(p,+1)=\left\vert p^-\right]$ &
$\bar{v}_\alpha(p,+1)=\left\langle p^-\right\vert$ \\
$v_\alpha(p,-1)=\left\vert p^-\right\rangle$ &
$\bar{v}_\alpha(p,-1)=\left[ p^-\right\vert$ \\
\end{tabular}
}
\caption{Assignment of quark and lepton wave functions. We label
the physical spin states by $j_3=\pm1$, which is twice the 3-component
of the spin. The wave functions assigned in Table~(a) are mapped onto
the bracket notation used in \SPINNEY{}~\cite{Cullen:2010jv} as defined
in Tables (b) and~(c).}\label{tab:wf-fermions}
\end{table*}

For spin~\nicefrac12 particles we use the assignment of wave functions
as shown in Table~\ref{tab:wf-fermions}; here, we quote the definition
of the massive spinors from~\cite{Cullen:2010jv} assuming the splitting
of Eq.~\eqref{eq:light-cone-splitting}:
\begin{subequations}
\begin{align}
\left\vert p^\pm\right\rangle&=\left\vert l\right\rangle
\pm\frac{\sqrt{p^2}}{\left[lk\right]}\left\vert k\right],\quad&
\left\vert p^\pm\right]&=\left\vert l\right]
\pm\frac{\sqrt{p^2}}{\left\langle lk\right\rangle}\left\vert k\right\rangle,\\
\left\langle p^\pm\right\vert&=\left\langle l\right\vert
\pm\frac{\sqrt{p^2}}{\left[kl\right]}\left[ k\right\vert,\quad&
\left[ p^\pm\right\vert&=\left[ l\right\vert
\pm\frac{\sqrt{p^2}}{\left\langle kl\right\rangle}\left\langle k\right\vert.
\end{align}
\end{subequations}
In order to preserve the condition that for any loop integral the
tensor rank does not exceed the number of loop propagators we fix
all gauge boson propagators to be in Feynman gauge. Their wave functions
are constructed as~\cite{Xu:1986xb}
\begin{equation}
\varepsilon_\mu(p,+1)=\frac{\left\langle q\vert\gamma_\mu\vert p^\flat\right]}{%
\sqrt{2}\left\langle qp^\flat\right\rangle},\quad
\varepsilon_\mu(p,-1)=\frac{\left[ q\vert\gamma_\mu\vert p^\flat\right\rangle}{%
\sqrt{2}\left[p^\flat q\right]},
\end{equation}
where $p^\flat=p$ in the massless case and $p^\flat=l$ according to
Eq.~\eqref{eq:light-cone-splitting} in the massive case. In the latter
case the third polarisation is defined as
\begin{equation}
\varepsilon_\mu(p, 0)=\frac1{\sqrt{p^2}}\left(2p_\mu^\flat-p_\mu\right).
\end{equation}
The wave functions and propagators for spin~\nicefrac{3}{2} and spin~2
particles correspond to those in~\cite{Kilian:2007gr}.

\paragraph{Simplifications}
Once all wave functions and propagators have been substituted by the
above definitions and all vertices have been replaced by their corresponding
expressions from the model file then all vector-like quantities and all
metric tensors are split into their four-dimensional and their orthogonal
part. As we use the 't~Hooft algebra, $\gamma_5$ is defined as a purely
four-dimensional object, $\gamma_5=i\epsilon_{\mu\nu\rho\sigma}%
\hat{\gamma}^\mu\hat{\gamma}^\nu\hat{\gamma}^\rho\hat{\gamma}^\sigma$.
By applying the usual anti-commutation relations for Dirac matrices we can
always separate the four-dimensional and $(n-4)$-dimensional parts of
Dirac traces, as we can use the fact that~\cite{Reiter:2009kb,Cullen:2010jv}
\begin{multline}
\mathrm{tr}(1)\cdot\mathrm{tr}(%
\hat{\gamma}_{\mu_1}\cdots\hat{\gamma}_{\mu_{l}}%
\tilde{\gamma}_{\mu_{l+1}}\cdots\tilde{\gamma}_{\mu_{l+p}})=\\
\mathrm{tr}(%
\hat{\gamma}_{\mu_1}\cdots\hat{\gamma}_{\mu_{l}}%
)\cdot\mathrm{tr}(%
\tilde{\gamma}_{\mu_{l+1}}\cdots\tilde{\gamma}_{\mu_{l+p}}).
\end{multline}
The same logic applies to open spinor lines such as~\cite{Cullen:2010jv}
\begin{multline}
\mathrm{tr}(1)\cdot\langle k_1\vert%
\hat{\gamma}_{\mu_1}\cdots\hat{\gamma}_{\mu_{l}}%
\tilde{\gamma}_{\mu_{l+1}}\cdots\tilde{\gamma}_{\mu_{l+p}}%
\vert k_2\rangle=\\
\langle k_1\vert %
\hat{\gamma}_{\mu_1}\cdots\hat{\gamma}_{\mu_{l}}%
\vert k_2\rangle\cdot\mathrm{tr}(%
\tilde{\gamma}_{\mu_{l+1}}\cdots\tilde{\gamma}_{\mu_{l+p}}).
\end{multline}
While the $(n-4)$-dimensional traces are reduced completely to
products of $(n-4)$-dimensional metric tensors $\tilde{g}^{\mu\nu}$,
the four-dimensional part is treated such that the number of terms in the
resulting expression is kept as small as possible. Any spinor line or trace
is broken up at any position where a light-like vector appears. Furthermore,
Chisholm idenities are used to resolve Lorentz contractions between
both Dirac traces and open spinor lines. If any traces remain we use
the built-in trace algorithm of \FORM{}~\cite{Vermaseren:2000nd}.

In the final result we can always avoid the explicit appearance of
Levi-Civit\'a tensors, noticing that any such tensor is contracted
with at least one light-like vector\footnote{Any external massive vector
at this point has been replaced by a pair of light-like ones. Contractions
between two Levi-Civit\'a symbols can be resolved to products of metric
tensors.}~$\hat{k}^\mu$,
and we can replace
\begin{equation}
\hat{k}^\mu\epsilon_{\mu\nu\rho\sigma}=-\frac{i}4\left(%
\left[k\vert\hat{\gamma}_\nu\hat{\gamma}_\rho\hat{\gamma}_\sigma%
      \vert k\right\rangle
-
\left\langle k\vert\hat{\gamma}_\nu\hat{\gamma}_\rho\hat{\gamma}_\sigma%
      \vert k\right]\right).
\end{equation}
Hence, the kinematic part of the numerator, at the end of our simplification
algorithm, is expressed entirely in terms of:
\begin{itemize}
\item spinor products of the form $\langle k_i k_j\rangle$, 
$[k_i k_j]$ or $[k_i\vert\hat{\gamma}^\mu\vert k_j\rangle\cdot\hat{q}_\mu$,
\item dot products $\hat{k}_i\cdot\hat{k}_j$
or $\hat{k}_i\cdot\hat{q}$,
\item constants of the Lagrangian such as masses, widths
      and coupling constants,
\item the symbols $\mu^2=\hat{q}^2-q^2$ and $\varepsilon=(n-4)/2$.
\end{itemize}

\paragraph{Treatment of $R_2$ rational terms}
In our representation for the numerator of one-loop diagrams, terms
containing the symbols $\mu^2$ or $\varepsilon$ can lead to a so-called
$R_2$ term~\cite{Ossola:2008xq}, which contributes to the rational part 
of the amplitude. In general, there are two ways of splitting
the numerator function:
\begin{subequations}
\begin{align}
\label{eq:r2:a}
\mathcal{N}(\hat{q},\mu^2,\varepsilon)&=
\mathcal{N}_0(\hat{q},\mu^2)+
\varepsilon\mathcal{N}_1(\hat{q},\mu^2)\nonumber\\
&\qquad+\varepsilon^2\mathcal{N}_2(\hat{q},\mu^2)\\
\intertext{or, alternatively,}
\label{eq:r2:b}
\mathcal{N}(\hat{q},\mu^2,\varepsilon)&=
\hat{\mathcal{N}}(\hat{q})+
\tilde{\mathcal{N}}(\hat{q},\mu^2,\varepsilon).
\end{align}
\end{subequations}
It should be noted that in Eq.~\eqref{eq:r2:a} the terms
$\mathcal{N}_1$ and $\mathcal{N}_2$ do not arise in DRED,
where only terms containing $\mu^2$ contribute to $R_2$.
Instead of relying on the construction of $R_2$ from specialized Feynman
rules~~\cite{Draggiotis:2009yb,Garzelli:2009is,Garzelli:2010qm,%
Garzelli:2010fq}, we generate the $R_2$ part along with all other
contributions without the need to separate the different parts.
For efficiency reasons, however, we provide an \emph{implicit} and
an \emph{explicit} construction of the $R_2$ terms. 

The implicit
construction uses the splitting of Eq.~\eqref{eq:r2:a} and treats
all three numerator functions~$\mathcal{N}_i$ on equal grounds.
Each of the three terms is reduced separately in a numerical reduction
and the Laurent series of the three results are added up taking into
account the powers of $\varepsilon$. 

The explicit construction of $R_2$
is based on the assumption that each term in
$\tilde{\mathcal{N}}$ in Eq.~\eqref{eq:r2:b} contains at least one
power of $\mu^2$ or $\varepsilon$. The expressions for those integrals
are relatively simple and known explicitly. Hence, the part of the amplitude
which originates from $\tilde{\mathcal{N}}$ is computed analytically whereas
the purely four-\hspace{0pt}dimensional part $\hat{\mathcal{N}}$ is passed to the numerical
reduction.

%% file: codegen.tex
\subsubsection{Abbreviation System}
To prepare the numerator functions of the one-loop diagrams for their numerical evaluation,
we separate the symbol $\mu^2$ and dot products involving the
momentum $\hat{q}$ from all other factors. All subexpressions which do
not depend on either $\hat{q}$ or $\mu^2$ are substituted by abbreviation
symbols, which are evaluated only once per phase space point.
Each of the two parts is then processed using \HAGGIES~\cite{Reiter:2009ts},
which generates optimized \FORTRAN{} code for their numerical evaluation.
For each diagram we generate an interface to
\SAMURAI~\cite{Mastrolia:2010nb}, \GOLEMVC~\cite{Cullen:2011kv} and/or
\PJFRY~\cite{Fleischer:2010sq}. The two latter codes are interfaced using
tensorial reconstruction at the integrand level~\cite{Heinrich:2010ax}.

\subsubsection{Reduction Strategies}
\label{ssec:reduction-strategies}
In the implementation of \GOSAM{}, great emphasis has been put on
maintaining flexibility with respect to the reduction algorithm that the user decides to use.
On the one hand, this is important because the best choice of the reduction method in terms of speed and numerical stability
can strongly depend on the specific process.
On the other hand, we tried to keep the code flexible to allow  further extensions to new
reduction libraries, such that \GOLEM{} can be used as a laboratory
for interfacing future methods with a realistic environment.

\begin{figure}
\centering
\includegraphics[width=\columnwidth]{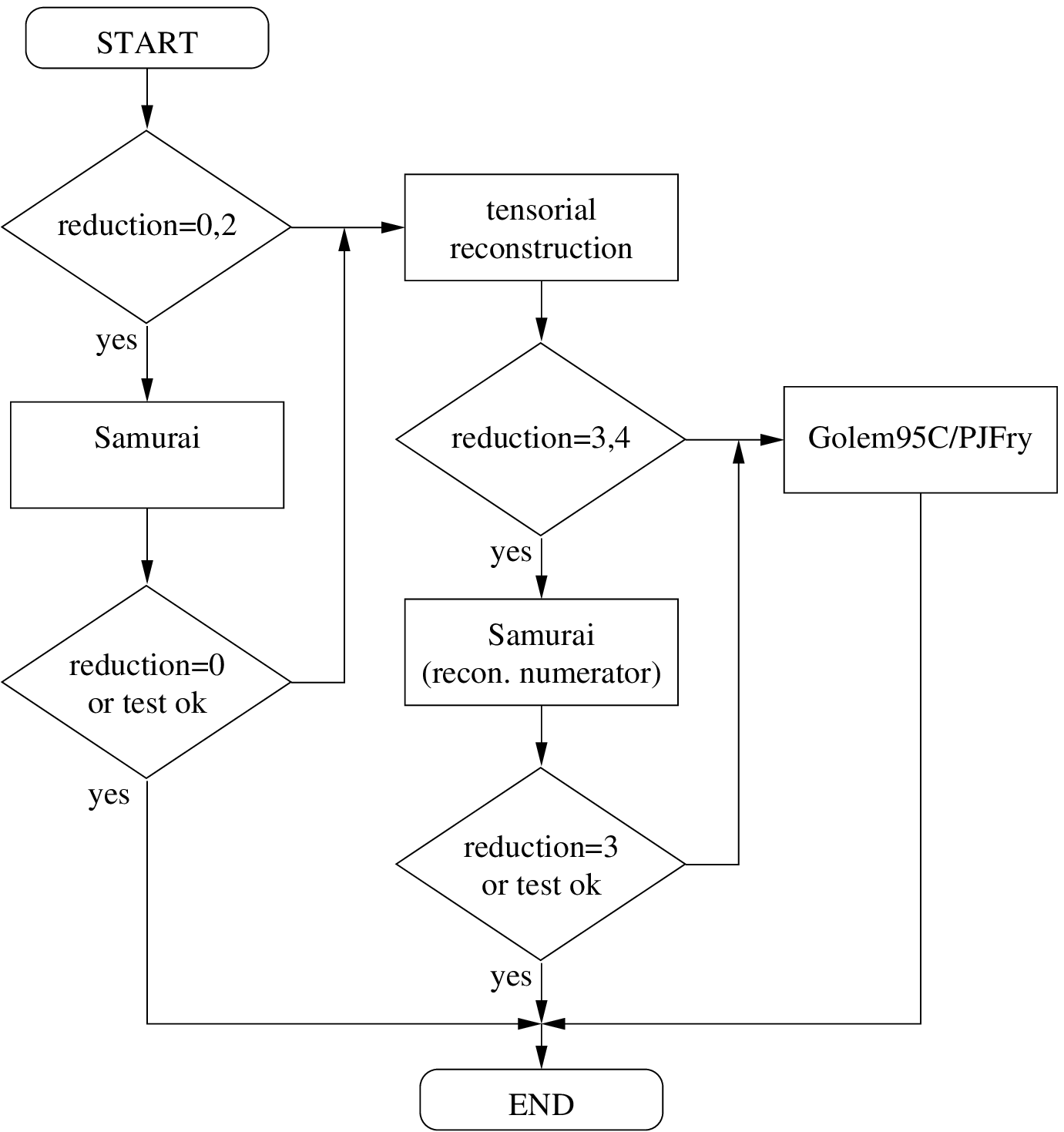}
\caption{Reduction strategies currently implemented in \GOLEM{}:
the reduction algorithm is chosen by  setting  the variable
\texttt{reduction\_interoperation} in the generated \FORTRAN{} code
and can be modified at run time. 0: \SAMURAI{} only;
1: \GOLEMVC{} only; 2: \SAMURAI{} with rescue option (\GOLEMVC);
3: \SAMURAI{} with numerator from tensorial reconstruction;
4: same as~3 but with rescue option(\GOLEMVC). 11, 12 and 14
are the same as 1, 2, 3 (respectively) with the difference
that \PJFRY{} is used instead of \GOLEMVC{}.}
\label{fig:reductionflow}
\end{figure}

Our standard choice for the reduction is \SAMURAI{}, which provides
a very fast and stable reduction in a large part of the phase space.
Furthermore, \SAMURAI{} reports to the client code if the quality
of the reconstruction of the numerator suffices the numerical requirements
(for details we refer to~\cite{Mastrolia:2010nb}). In \GOLEM{} we use this
information to trigger an alternative reduction with either
\GOLEMVC~\cite{Cullen:2011kv} or \PJFRY~\cite{Fleischer:2010sq} whenever
these reconstruction tests fail, as shown in Fig.~\ref{fig:reductionflow}.
The reduction algorithms implemented in these libraries extend to phase space
regions of small Gram determinants and therefore cover most cases in which
on-shell methods cannot operate sufficiently well.
This combination of on-shell techniques and traditional tensor reduction
is achieved using tensorial reconstruction at the integrand
level~\cite{Heinrich:2010ax}, which also provides the possibility of running
on-shell methods with a reconstructed numerator. 
In addition to solving the problem of numerical instabilities, in some cases this option 
can reduce the computational cost of the reduction.
Since the reconstructed numerator is typically of a form where 
kinematics and loop momentum dependence are already separated , 
the use of a reconstructed numerator tends to be faster than the original procedure, 
in particular in cases with a large number of legs and low rank. 

The flowchart in  Fig.~\ref{fig:reductionflow} summarizes all possible reduction strategies
which are currently implemented. The strategy in use is selected by
assigning the variable \texttt{reduction\_interoperation} in the generated
\FORTRAN{} code. The availability of the branches is determined during
code generation by activating (at least one of) the extensions
(\texttt{samurai}, \texttt{golem95}, \texttt{pjfry}) 
in the input card. Switching between
active branches is possible at run time. In detail, the possible choices for the variable \texttt{reduction\_interoperation}
are the following: 
\begin{description}
\item[0] the numerators of the one-loop diagrams are reduced by \SAMURAI{},
         no rescue system is used in case the reconstruction test fails;
\item[1] the tensor coefficients of the numerators are reconstructed
         using the \emph{tensorial reconstruction at the integrand level},
         the numerator is expressed in terms of tensor integral form factors
         which are evaluated using \GOLEMVC{};
\item[2] the numerators are reduced by \SAMURAI{}; whenever the
         reconstruction test fails, numerators are reduced using the option~1 as a
         backup method;
\item[3] tensorial reconstruction is used to compute the tensor coefficients;
         \SAMURAI{} is employed for the reduction of the reconstructed numerator, no rescue system is used;
\item[4] as in option~3, \SAMURAI{} is used to reduce the reconstructed
         numerator, \GOLEMVC{} is used as backup option;
\item[11] same as~1 but \PJFRY{} is used instead of \GOLEMVC{};
\item[12] same as~2 but \PJFRY{} is used instead of \GOLEMVC{};
\item[14] same as~4 but \PJFRY{} is used instead of \GOLEMVC{}.
\end{description}

It is difficult to make a statement about the ``optimal" reduction method because this 
depends on the process under consideration. For multi-leg processes, e.g. 
$b\bar{b}b\bar{b}$ production, we found that \SAMURAI{} 
is clearly superior to tensor reduction in what concerns timings and size of the code. 
Concerning points which need a special treatment, we did not make extensive studies 
using traditional tensor reduction only, but one can certainly say that 
the combination of \SAMURAI{} and tensorial 
reconstruction seems to be optimal in what concerns the avoidance of numerical 
instabilities due to inverse Gram determinants.


%% file: conventions.tex
In this section we briefly discuss the conventions chosen for the results
returned by \GOLEM{}. 
Depending on the actual setup for a given
process, in particular if an imported model file is used, conventions may be slightly different. 
Here we restrict the discussion to the
case where the user wants to compute QCD corrections to a process and
in the setup files he has put $g_s=1$. In this case, the tree-level
matrix element squared can be written as
\begin{equation}\label{eq:amp0:def}
\vert\mathcal{M}\vert_{\text{tree}}^2=\mathcal{A}_0^\dagger\mathcal{A}_0=
(g_s)^{2b}\cdot a_0\;.
\end{equation}
The fully renormalised matrix element at one-loop level, i.e. the
interference term between tree-level and one-loop, can be written as
\begin{multline}\label{eq:amp1:def}
\vert\mathcal{M}\vert_{\text{1-loop}}^2=
\mathcal{A}_1^\dagger\mathcal{A}_0+
\mathcal{A}_0^\dagger\mathcal{A}_1=
2\cdot\Re(\mathcal{A}_0^\dagger\mathcal{A}_1)=
\\
\vert\mathcal{M}\vert^2_{\text{bare}}
+\vert\mathcal{M}\vert^2_{\text{ct, $\delta m_Q$}}
+\vert\mathcal{M}\vert^2_{\text{ct, $\alpha_s$}}
+\vert\mathcal{M}\vert^2_{\text{wf, g}}
+\vert\mathcal{M}\vert^2_{\text{wf, Q}}=\\
\frac{\alpha_s(\mu)}{2\pi}\frac{(4\pi)^\varepsilon}{\Gamma(1-\varepsilon)}
\cdot (g_s)^{2b}\cdot\left[%
c_0+\frac{c_{-1}}{\varepsilon}+\frac{c_{-2}}{\varepsilon^2}
+\mathcal{O}(\varepsilon)%
\right]\;.
\end{multline}
A call to the subroutine \texttt{samplitude} returns an array
consisting of the four numbers $(a_0, c_0, c_{-1}, c_{-2})$
in this order. 
The average over initial state colours and helicities is included 
in the default setup. 
In cases where the process is loop induced, i.e. the tree level amplitude is absent, 
the program returns the values for $\mathcal{A}_1^\dagger\mathcal{A}_1$ 
where a factor $$\left(\frac{\alpha_s(\mu)}{2\pi}\frac{(4\pi)^\varepsilon}{\Gamma(1-\varepsilon)}\right)^2$$ has been pulled out.

After all UV-renormalisation contributions have been taken
into account correctly, only IR-singularities remain, which can be
computed using the routine \texttt{ir\_sub\-trac\-tions}. This routine
returns a vector of length two, containing the coefficients of the
single and the double pole, which should be equal to $(c_{-1}, c_{-2})$
and therefore can be used as a check of the result.

\paragraph{Ultraviolet Renormalisation in QCD}
For UV-\hspace{0pt}renormalisation we use the $\overline{\text{MS}}$ scheme for
the gluon and all massless quarks, whereas a subtraction at zero momentum
is chosen for massive quarks~\cite{Nason:1987xz}. Currently, counterterms
are only provided for QCD corrections. In the case of electroweak corrections
only unrenormalised results can be produced automatically.

\begin{figure}[h]
\centering
\includegraphics[scale=0.8]{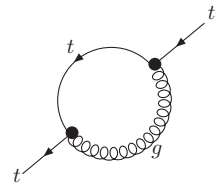}
\caption{\label{fig:algebra:mqse} Feynman diagram of a massive quark
self energy in QCD. For this type of diagram \GOLEM{} automatically
generates UV-counterterms.}
\end{figure}

For computations involving loop propagators for massive fermions,
we introduced the automatic generation of a mass counter term needed
for the on-shell renormalisation of the massive particle.
Here, we exploit the fact
that such a counter term is strictly related to the massive fermion
self energy bubble diagrams (see Fig.~\ref{fig:algebra:mqse}). 
As described in Section~\ref{ssec:diagen}, the program \GOLEM{} analyzes
all generated diagrams. In that step also self-energy insertions of massive
quarks are detected, where we make the replacement
\begin{multline}
\frac{(\fmslash{q}+\fmslash{r}+m)\cdot g^{\mu\nu}}{%
\left[(q+r)^2-m^2\right]q^2} \to
\frac{(\fmslash{q}+\fmslash{r}+m)\cdot g^{\mu\nu}}{%
\left[(q+r)^2-m^2\right]q^2}
\\+\frac{m}{4}\left[
\frac{6q\cdot r+3(r^2-m^2)}{m^2}
\right.\\\left.+\frac{3(4+1_{\mathrm{HV}})\mu^2}{r^2-3m^2}
\right]%
\frac{g^{\mu\nu}}{%
\left[(q+r)^2-m^2\right]q^2}.
\end{multline}
The symbol $1_{\mathrm{HV}}$ is one in the \tHV{} scheme and
zero in DRED. 

Performing the integral, contracting the expression with the QCD vertices at both sides
and multiplying the missing factor of $(2\pi)^{-1}$ we retrieve the
expression for the mass counter-term,
\begin{equation}
\label{eq:mass-ct:final}
\frac{\delta m}{m}= \frac{\alpha_s}{2\pi}\frac{(4\pi)^\varepsilon}{\Gamma(1-\varepsilon)}
\frac{C_F}{2}\left(\frac{\mu^2}{m^2}\right)^\varepsilon%
\left[\frac{3}{\varepsilon}+5-1_{\mathrm{HV}}\right].
\end{equation}

Furthermore, the renormalisation of $\alpha_s$ leads to a term of the form
\begin{multline}\label{eq:uv-delta:as}
\vert\mathcal{M}\vert^2_{\text{ct, $\alpha_s$}}=b\cdot\frac{\alpha_s}{2\pi}
\frac{(4\pi)^\varepsilon}{\Gamma(1-\varepsilon)}
\vert\mathcal{M}\vert^2_{\text{tree}}\cdot
\left[%
-\frac{\beta_0}{\varepsilon}
\right.\\\left.
+\frac{2T_R}{3\varepsilon}%
\sum_{q=N_f+1}^{N_f+N_{f,h}}\left(\frac{\mu^2}{m_q^2}\right)^\varepsilon
+\frac{C_A}{6}(1-1_{\mathrm{HV}})%
\right],
\end{multline}
with $\beta_0=(11C_A-4T_RN_f)/6$, $N_f$ being the number of light
quark flavours, $N_{f,h}$ the number of heavy flavours, 
and $b$ is the power of the coupling in the  Born amplitude as defined in Eq.~\eqref{eq:amp0:def}.
The last term of Eq.~\eqref{eq:uv-delta:as}
provides the finite renormalisation needed to compensate
the scheme dependence of $\alpha_s$,
\begin{equation}
\alpha_s^{\mathrm{DR}}=\alpha_s^{\overline{\mathrm{MS}}}\left(%
1+\frac{C_A}{6}\frac{\alpha_s^{\overline{\mathrm{MS}}}}{2\pi}\right).
\end{equation}
A further contribution consists of the wave-function renormalisation
of massive external quark lines. If we denote the set of external massive
quark lines by $\mathcal{Q}_h=\{Q_1(m_1),\ldots,Q_p(m_p)\}$ we obtain
\begin{multline}
\vert\mathcal{M}\vert^2_{\text{wf, Q}}=-\frac{\alpha_s}{2\pi}
\frac{(4\pi)^\varepsilon}{\Gamma(1-\varepsilon)}\frac{C_F}{2}%
\times\\
\sum_{Q(m)\in\mathcal{Q}_h}%
\left(\frac{\mu^2}{m^2}\right)^\varepsilon\left[\frac3{\varepsilon}%
+5-1_{\mathrm{HV}}
\right]\cdot\vert\mathcal{M}\vert^2_{\text{tree}},
\end{multline}
Finally, also the wave function of the gluon receives a contribution
from the presence of heavy quarks in closed fermion loops. If $N_g$
is the number of external gluon lines, this contribution can be written
as
\begin{multline}\label{eq:uv-delta:wfg}
\vert\mathcal{M}\vert^2_{\text{wf, g}}=-\frac{\alpha_s}{2\pi}
\frac{(4\pi)^\varepsilon}{\Gamma(1-\varepsilon)}\,N_g\,\frac{2T_R}{3\varepsilon}%
\times\\
\sum_{q=N_f+1}^{N_f+N_{f,h}}\left(\frac{\mu^2}{m_q^2}\right)^\varepsilon
\cdot\vert\mathcal{M}\vert^2_{\text{tree}},
\end{multline}

At the level of the generated \FORTRAN{} code the presence of these
contributions can be controlled by a set of variables defined in the
module \texttt{config.f90}. The variable \texttt{renormalisation}
can be set to \texttt{0}, \texttt{1}, or~\texttt{2}.
If \texttt{renormalisation=0}, none of the counterterms ar present. If
\texttt{renormalisation=2} only
$\vert\mathcal{M}\vert^2_{\text{ct, $\delta m_Q$}}$ is included, 
which is the counter\-term stemming from all terms of the type of Eq.~\eqref{eq:mass-ct:final}
contributing to the amplitude.\\
In the case where \texttt{renormalisation=1} a more fine-grained control
over the counterterms is possible.
\begin{description}
\item{\texttt{renorm\_logs}:} if set to \emph{false}, in all counterterms
the generation of logarithms is disabled, i.e. factors of the form
$(\bullet)^\varepsilon$ in eqs.~\eqref{eq:mass-ct:final} to 
\eqref{eq:uv-delta:wfg} are replaced by one.
\item{\texttt{renorm\_beta}:} if set to \emph{false}, the counterterm
$\vert\mathcal{M}\vert^2_{\text{ct, $\alpha_s$}}$ is set to zero.
\item{\texttt{renorm\_mqwf}:} if set to \emph{false}, the counterterm
$\vert\mathcal{M}\vert^2_{\text{wf, Q}}$ is set to zero.
\item{\texttt{renorm\_mqse}:} if set to \emph{false}, the counterterm
$\vert\mathcal{M}\vert^2_{\text{ct, $\delta m_Q$}}$ is set to zero.
\item{\texttt{renorm\_decoupling}} if set to \emph{false}, the counterterm
$\vert\mathcal{M}\vert^2_{\text{wf, g}}$ is set to zero.
\end{description}
The default settings for \texttt{renormalisation=1} are \emph{true}
for all the \texttt{renorm} options listed above.

\paragraph{Finite Renormalisation of $\gamma_5$ in QCD}
In the \tHV{} scheme, a finite renormalisation term for $\gamma_5$
is required beyond tree level. The relevant terms are generated only if
\texttt{fr5} is added in the input card to the list of extensions before code generation.
Currently, the automatic generation of this finite contribution is not
performed if model files different from the built-in model files are used. 
In agreement
with~\cite{Harris:2002md} and~\cite{Weinzierl:1999xb} we replace
the axial component at each vertex,
\begin{align}\label{eq:renorm:fr5}
\gamma^\mu\gamma_5&\to
\frac12Z_{\text{axial}}%
\left(\gamma^\mu\gamma_5-\gamma_5\gamma^\mu\right),\\
\intertext{with}
Z_{\text{axial}}&=1-2\frac{\alpha_s}{2\pi}C_F\cdot 1_{\mathrm{HV}}.
\end{align}
Once it is generated, this contribution can be switched on and off at run-time
through the variable \texttt{re\-norm\_gam\-ma5}, which is defined
in the module \texttt{config.f90}.

\paragraph{Conversion between the Schemes}
In \GOLEM{} we have implemented two different schemes,
the \tHV{} scheme and dimensional reduction.
By default, the former is used, while the latter can be activated by
adding the extension \texttt{dred}. If a QCD computation has been done in
dimensional reduction the result can be converted back to the  \tHV{}
scheme by adding a contribution for each external massless parton,
\begin{multline}
\vert\mathcal{M}^{\text{CDR}}\vert_{\text{1-loop}}^2=
\vert\mathcal{M}^{\text{DR}}\vert_{\text{1-loop}}^2\\
-\frac{\alpha_s}{2\pi}\vert\mathcal{M}^{\text{DR}}\vert_{\text{tree}}^2
\sum_{I=1}^{N_{\text{ext}}}\tilde{\gamma}_I^{\text{DR}},
\end{multline}
with $\tilde{\gamma}_q^{\text{DR}}=\tilde{\gamma}_{\bar{q}}^{\text{DR}}=C_F/2$
and $\tilde{\gamma}_g^{\text{DR}}=C_A/6$.
This conversion can be switched on by setting \texttt{convert\_to\_cdr} to \emph{true} 
in the module \texttt{config.f90}.
At one-loop level, the \tHV{} scheme and conventional dimensional
regularisation~(CDR) are equivalent in the sense that
$\tilde{\gamma}_I^{\text{'t\,HV}}=0$ for all partons.

%% file: install.tex
\subsection{Requirements}
\noindent
The program \GOLEM{} is designed to run in any modern Linux/Unix environment;
we expect that \PYTHON~($\geq2.6$), Java~($\geq1.5$) and Make are installed
on the system. Furthermore, a \FORTRAN~95 compiler is required in order to
compile the generated code. Some \FORTRAN~2003 features are used if one
wants to make use of the Les Houches interface~\cite{Binoth:2010xt}.
We have tried all examples using \texttt{gfortran} versions 4{.}1 and 4{.}5.

On top of a standard Linux environment, the programs
\FORM~\cite{Vermaseren:2000nd}, 
version~$\geq3.3$ (newer than Aug.11,~2010) and
\QGRAF~\cite{Nogueira:1991ex} need to be installed on the system.
Where\-as \SPINNEY{}~\cite{Cullen:2010jv} and \HAGGIES{}~\cite{Reiter:2009ts}
are part of \GOLEM{} and are not required to be installed separately,
at least one of the libraries
\SAMURAI{}~\cite{Mastrolia:2010nb} and \GOLEMVC{}~\cite{Cullen:2011kv}
needs to be present at compile time of the generated code. Optionally,
\PJFRY{}~\cite{Fleischer:2010sq} can be used on top of \GOLEMVC{}.

\subsection{Download and Installation}
\paragraph{\QGRAF} The program can be downloaded as \FORTRAN{}
    source code from 
        \bcen
    \url{http://cfif.ist.utl.pt/~paulo/qgraf.html}\, .
 \ecen
    After unpacking the tar-ball,
    a single \FORTRAN{}\,\texttt{77} file needs to be compiled.
\paragraph{\FORM} 
    The program is available
    at 
    \bcen
    \url{http://www.nikhef.nl/~form/} 
    \ecen
  both as a compiled binary
    for many platforms and as a tar-ball. The build process,
    if built from the source files, is controlled by \texttt{Autotools}.
\paragraph{\SAMURAI{} and \GOLEMVC} These libraries are available as tar-balls
    and from subversion repositories
    at
    \bcen
\url{http://projects.hepforge.org/samurai/}
	\ecen
	 and
	 \bcen
\url{http://projects.hepforge.org/golem/95/}\ecen
 respectively.
    For the user's convenience we have prepared a package containing
    \SAMURAI{} and \GOLEMVC{} together with the integral libraries
    \texttt{One\-LOop}~\cite{vanHameren:2010cp},
    \texttt{QCD\-Loop}~\cite{Ellis:2007qk} and
    \texttt{FF}~\cite{vanOldenborgh:1989wn}.
   The package \texttt{gosam-\hspace{0pt}contrib-\hspace{0pt}1.0.tar.gz}
   containing all these libraries is available
   for download from: 
\bcen   
   \url{http://projects.hepforge.org/gosam/}
\ecen
\paragraph{\GOLEM} The user can download the code either as a tar-ball
    or from the subversion repository
    at
    \bcen
    \url{http://projects.hepforge.org/gosam/}\, .
	\ecen
    The build process and in\-stal\-la\-tion of \GOLEM{} is controlled by
    \PYTHON{} \texttt{Dist\-utils}, while the build process for the libraries 
    \SAMURAI{} and \GOLEMVC{}
    is controlled by \texttt{Autotools}.

\noindent Therefore the installation proceeds in two steps:
\begin{enumerate}
\item For all components which use \texttt{Autotools}, the following
sequence of commands installs them under the user defined directory
\texttt{MYPATH}.
\begin{lstlisting}
./configure --prefix=MYPATH
make FC=gfortran F77=gfortran
make install # or sudo make install
\end{lstlisting}
If the \texttt{configure} script is not present, the user needs to run
\texttt{sh ./autogen.sh} first.
\item For \GOLEM{} which is built using \texttt{Distutils}, the user needs to run
\begin{lstlisting}
python setup.py install \
     --prefix MYPATH
\end{lstlisting}
If \texttt{MYPATH} is different from the system default (e.g. {\it /usr/bin}), 
the environment
variables \texttt{PATH}, \texttt{LD\_LIB\-RA\-RY\_PATH} and \texttt{PYTHONPATH}
might have to be set accordingly. For more details we direct the user to
the \GOLEM{} reference manual and to the documentation of the beforementioned
programs.
\end{enumerate}

%% file: usage.tex
\GOLEM{} is a very flexible program and comes with a wide range of
configuration options. Not all of these options are relevant for
simple processes and often the user can leave most of the settings at their
default values. In order to generate the code for a process,
one needs to prepare an input file, which will be
called \textit{process card} in the following, which contains
\begin{itemize}
\item process specific information, such as a list of initial and
      final state particles, their helicities (optional) 
      and the order of the coupling constants;
\item scheme specific information and approximations, such as
      the regularisation and renormalisation schemes,
      the underlying model,
      masses and widths which are set to zero, 
      the selection of subsets of diagrams; the latter might
      be process dependent;
\item system specific information, such as paths to programs and libraries
      or compiler options;
\item optional information for optimisations which control the code generation.
\end{itemize}
In the following we explain how to set up the required files for the
process $q\bar{q}\to gZ^0\to g\,e^-e^+$. The example
computes the QCD corrections for the $u\bar{u}$ initial state, where $m_e=0$
and $N_f=5$ massless quarks are assumed.
For our example, 
we follow an approach where we keep the different types of information in
separate files -- {\tt process.rc}, {\tt scheme.rc} and {\tt system.rc} --
and use \GOLEM{} to produce a process card for this process based on
these files.
This is not required --- one could also produce and edit the process card
directly --- it is however more convenient to 
store system specific information into a separate, re-usable file,
and it makes the code generation more transparent.
 
\paragraph{Process specific information}
The following listing contains the information which is specific to
the process. The syntax of process cards requires that no blank character
is left between the equals sign and the property name. Commentary can be added
to any line, marked by the `\texttt{\#}' character. Line
continuation is achieved using a backslash at the end of a line.\footnote{%
The line numbers are just for reference and should not be included in the actual
files.}
\begin{lstlisting}[numbers=left,caption={File '\texttt{process.rc}'},%
label=lst:process.rc]
process_path=qqgz
in=u,u~
out=g,e-,e+
helicities=+-+-+,+---+,-++-+,-+--+
order=QCD,1,3
\end{lstlisting}
The first line defines the (relative) path to the directory where the
process files will be generated. \GOLEM{} expects that this directory
has already been created.
Lines 2 and~3 define the initial and final state of the process in
terms of field names, which are defined in the model file. Besides
the field names one can also use PDG codes~\cite{Yost:1988ke,Caso:1998tx}
instead. Hence, the following lines would be equivalent to lines 2 and~3
in Listing~\ref{lst:process.rc}:
\begin{lstlisting}[numbers=left,firstnumber=2]
in=2,-2
out=21,11,-11
\end{lstlisting}

Line~4 describes the helicity amplitudes which should be generated.
If no helicities are specified, the program defaults to the generation of 
all possible helicity configurations, some of which may turn out to be zero.
The different helicity amplitudes are separated by commas; within
one helicity amplitude there is one character (usually `\texttt{+}',
`\texttt{-}' and \texttt{`0'}) per external particle from the left to the
right. In the above example for the reaction
\begin{displaymath}
u(k_1,\lambda_1)\bar{u}(k_2,\lambda_2)\to
g(k_3,\lambda_3)e^-(k_4,\lambda_4)e^+(k_5,\lambda_5)
\end{displaymath}
we have the following assignments:
\begin{center}
\begin{tabular}{l|lllll}
Helicity & $\lambda_1$&$\lambda_2$&$\lambda_3$&$\lambda_4$&$\lambda_5$\\
\hline
0&+&-&+&-&+ \\
1&+&-&-&-&+ \\
2&-&+&+&-&+ \\
3&-&+&-&-&+
\end{tabular}
\end{center}
With the above value for \texttt{helicities} we generate all non-vanishing
helicities for the partons but keep the lepton helicities fixed. In more
complicated examples this way of listing all helicities explicitly can be
very tedious.
Therefore, we introduced the option to generate sets of helicities using
square brackets.
For example, if the gluon helicity is replaced by \texttt{[+-]}, 
the bracket is expanded automatically to take the values {\tt +,-}.
\begin{lstlisting}[numbers=left,firstnumber=4]
helicities=+-[+-]-+, -+[+-]-+
\end{lstlisting}
A further syntactical reduction can be achieved for the quarks.
The current expansion of a square bracket and its opposite value
can be assigned to a pair of variables as in \texttt{[xy=+-]}.
If the bracket expands to `\texttt{+}' then \texttt{x} is assigned
`\texttt{+}' and \texttt{y} is assigned the opposite sign, i.e. `\texttt{-}'.
If the bracket expands to `\texttt{-}' the assignments are
\texttt{x=-} and \texttt{y=+}. Hence, the helicity states of a massless
quark anti-quark pair are generated by \texttt{[qQ=+-]Q},
and the selection of helicities in our example can be abbreviated to
\begin{lstlisting}[numbers=left,firstnumber=4]
helicities=[qQ=+-]Q[+-]-+
\end{lstlisting}
which is equivalent to the version of this line in
Listing~\ref{lst:process.rc}.

Finally, the order (power) of the coupling constants has to be specified.
Line~5 contains  a keyword for the type of coupling 
(\texttt{QCD} or \texttt{QED}),
the order of this coupling constant in the unsquared tree level amplitude
(in our example:~1)
and the order of the coupling constant in the unsquared one-loop amplitude
(in our example:~3). One can also use \GOLEM{} to generate the tree level only,
by giving only the power of the tree level amplitude:
\begin{lstlisting}[numbers=left,firstnumber=5]
order=QCD,1
\end{lstlisting}
Conversely, \GOLEM{} will generate the virtual amplitude squared for
processes where no tree level is present if the tree level order
is replaced by the keyword \texttt{NONE}.
\begin{lstlisting}[numbers=left,firstnumber=5]
order=QCD,NONE,3
\end{lstlisting}

Up to now, the file would generate all 8 tree level and 180 one-loop diagrams
contributing to the process $u\bar{u}\to g\, e^-e^+$, regardless of the
intermediate states. Nevertheless, what we intended to generate were only
those diagrams where the electron pair comes from the decay of
a $Z\to e^-e^+$. \GOLEM{} offers two ways of achieving this diagram
selection, either by passing a condition to \QGRAF{} or by applying a
filter written in \PYTHON. The first option would be specified by
the option \texttt{qgraf.}\hspace{0pt}\texttt{verbatim},
which copies the argument of the
option to the \QGRAF{} input file in verbatim. The following filter
demands the appearance of exactly one $Z$-propagator, leaving us
with 2 tree-level and 45 one-loop diagrams:
\begin{lstlisting}[numbers=left,firstnumber=6]
qgraf.verbatim= true=iprop[Z,1,1];
\end{lstlisting}
The alternative solution is the application of a \PYTHON{} filter
using the options \texttt{filter.lo} for tree level and
\texttt{filter.nlo} for one-loop diagrams. The current example requires
the two lines
\begin{lstlisting}[numbers=left,firstnumber=6]
filter.lo=   IPROP([Z])==1
filter.nlo=  IPROP([Z])==1
\end{lstlisting}

\paragraph{Scheme specific information}
For our example we put all scheme specific definitions in the file
\texttt{scheme.rc}. It contains the choice of a suitable regularisation
scheme and fixes what types of UV counterterms are included in the final result.
\begin{lstlisting}[numbers=left,caption={File '\texttt{scheme.rc}'},%
label=lst:scheme.rc]
extensions=dred
qgraf.options=onshell
zero=mU,mD,mC,mS,mB,me,wT
one=gs
\end{lstlisting}

In Listing~\ref{lst:scheme.rc}, line~1 selects dimensional reduction
as a regularisation scheme. If \texttt{dred} is not specified in the
list of extensions, \GOLEM{} works in the \tHV{} scheme by
default. Line~2 removes all on-shell bubbles on external legs. This
is, on the one hand, required to be consistent with the renormalisation
scheme. On the other hand, those diagrams would lead to zero denominators
at the algebraic level. In line~3 all light quark masses, the mass of the electron 
and the width of the top quark are set to zero.
Further, as a convention rather than a scheme, 
the strong coupling $g_s$ is set to one in line~4, which means that $g_s$
will not occur in the algebraic expressions, assuming that the user will 
multiply the final result by his desired value for the strong coupling. 
If the option {\tt one=gs} is not used, the default value contained in 
the file {\tt common/model.f90} will be used. This default value of course 
can be changed by the user.

\paragraph{System specific information}
In order to adapt the code generation to the system environment,
\GOLEM{} needs to find a way of determining all relevant paths and
options for the programs and libraries used during generation, compilation
and linking of the code. Those settings are fixed in the file
\texttt{system.rc} in our example.\footnote{In this example we
assume that the user has defined an environment variable
\texttt{PREFIX}.}
\begin{lstlisting}[numbers=left,caption={File '\texttt{system.rc}'},%
label=lst:system.rc]
system.extensions=samurai,golem95
samurai.fcflags=\
   -I${PREFIX}/include/samurai
samurai.ldflags=\
   -L${PREFIX}/lib -lsamurai
samurai.version=2.1.1
golem95.fcflags=\
   -I${PREFIX}/include/golem95
golem95.ldflags=\
   -L${PREFIX}/lib -lgolem95
form.bin=${PREFIX}/bin/tform
qgraf.bin=${PREFIX}/bin/qgraf
fc.bin=gfortran
\end{lstlisting}

\paragraph{Generating the Code}
After having prepared the input files correctly we need to collect the
information distributed over the three files \texttt{process.rc},
\texttt{scheme.rc} and \texttt{system.rc} in one input file, which we
will call \texttt{gosam.in} here. The corresponding command is:
\begin{lstlisting}
gosam.py --template gosam.in \
  --merge process.rc \
  --merge scheme.rc --merge system.rc
\end{lstlisting}
The generated file can be processed with \texttt{gosam.py} directly
but requires the process directory to be present.
\begin{lstlisting}
mkdir qqgz
gosam.py gosam.in
cd qqgz
\end{lstlisting}
All further steps are controlled by the generated make files;
in order to generate and compile all files relevant for the matrix element
one needs to invoke
\begin{lstlisting}
make compile
\end{lstlisting}
The generated code can be tested with the program \texttt{matrix/test.f90}.
The following sequence of commands will compile and run the example program.
\begin{lstlisting}
cd matrix
make test.exe
./test.exe
\end{lstlisting}
The last lines of the program output should look as follows\footnote{%
The actual numbers depend on the random number generator of the system
because the phase space point is generated randomly;
however, the pole parts should agree between the matrix element and
the infrared insertion operator given that the matrix element is fully
renormalised.}
\begin{lstlisting}
#             LO:  0.3450350717601E-06
# NLO, finite part  -10.77604823456547
# NLO, single pole  -19.98478948141949
# NLO, double pole  -5.666666665861926
# IR,  single pole  -19.98478948439310
# IR,  double pole  -5.666666666666666
\end{lstlisting}
The printed numbers are, in this order, $a_0$, $c_0/a_0$, $c_{-1}/a_0$,
$c_{-2}/a_0$ and the pole parts calculated from the infrared insertion
operator~\cite{Catani:1996vz,Catani:2000ef}.

One can generate a visual representation of all generated diagrams using
the command
\begin{lstlisting}
make doc
\end{lstlisting}
which generates the file \texttt{doc/process.ps} using a \PYTHON{}
implementation of the algorithm described in~\cite{Ohl:1995kr} and
the \LaTeX{} package AXODRAW~\cite{Vermaseren:1994je}.

\subsubsection{Further Options}
\GOLEM{} provides a range of options which influence the code generation,
the compilation and the numerical evaluation of the amplitude.
Giving an exhaustive list of all options would be far beyond the scope of
this article and the interested user is referred to the reference manual.
Nonetheless, we would like to point out some of \GOLEM{}'s capabilities
by presenting the corresponding options.

\paragraph{Generating the $R_2$ Term}
When setting up a process the user can specify if and how the
$R_2$ term of the amplitude should be generated by setting the variable
\texttt{r2} in the setup file.
\begin{lstlisting}
r2=explicit
\end{lstlisting}
Possible options for \texttt{r2} are \texttt{implicit}, which is
the default, \texttt{explicit}, \texttt{off} and \texttt{only}.
The keyword \texttt{implicit} means that the $R_2$ term is
generated along with the four-dimensional numerator
as a function in terms of $\hat{q}$, $\mu^2$ and $\varepsilon$ and
is reduced at runtime by sampling different values for~$\mu^2$.
This is the slowest but also the most general option. Using the keyword
\texttt{explicit} carries out the reduction of terms containing
$\mu^2$ or $\varepsilon$ during code generation
(see Appendix~\ref{ssec:app-r2}). The keyword \texttt{off} puts the $R_2$
term to zero which is useful if the user wants to provide his own calculation
for these terms. Conversely, using \texttt{r2=only} discards everything but
the $R_2$ term (reducing it as in the case \texttt{explicit})
and puts \GOLEM{} in the position of providing $R_2$ terms
for external codes which work entirely in four dimensions.

\paragraph{Diagram Selection}
\GOLEM{} offers a two-fold way of selecting and discarding diagrams.
One can either influence the way \QGRAF{} generates diagrams or
apply filters to the diagrams after they have been generated by \QGRAF{}
or combine the two methods. Let us assume that in the above example we
want to remove the third generation of quarks completely. Hence, all closed
quark loops would be massless and therefore the second generation is
just an exact copy of the first one. We can therefore restrict the generation
of closed quark loops to up and down quarks.
\GOLEM{} has a filter precisely for this purpose, which takes the
field names of the flavours to be generated as arguments.
\begin{lstlisting}
filter.nlo=NFGEN(U,D)
\end{lstlisting}
This filter can be combined with the already existing filter selecting
only diagrams containing a $Z$-propagator using the \texttt{AND} function:
\begin{lstlisting}
filter.nlo=AND( NFGEN(U,D), \
              IPROP([Z]) == 1 )
\end{lstlisting}

A further feature of the code generated by \GOLEM{} is the possibility
of selecting diagrams at runtime. For example, we would like to distinguish
at runtime three different gauge invariant sets of  diagrams at one-loop level:
\begin{enumerate}
\item diagrams with a closed quark loop where the $Z$ is attached to the loop;
\item diagrams with a closed quark loop where the $Z$ is emitted from the
      external quark line;
\item diagrams without a closed quark loop.
\end{enumerate}
In order to provide the code for a diagram selection at runtime one simply
replaces the above filter by a list of filters as follows
\begin{lstlisting}
filter.nlo=[\
  AND( NFGEN(U,D), IPROP([Z]) == 1,  \
    NF, LOOPVERTICES([Z],_,_) == 1), \
  AND( NFGEN(U,D), IPROP([Z]) == 1,  \
    NF, LOOPVERTICES([Z],_,_) == 0), \
  AND( NFGEN(U,D), IPROP([Z]) == 1,  \
    NOT(NF))]
\end{lstlisting}
The two new filters in use are \texttt{NF} which selects closed quark loops
only and \texttt{LOOPVERTICES} which counts the number of vertices
attached to the loop with the given sets of fields running through the vertex
(where \texttt{\_} replaces any field). In the \FORTRAN{} files one can
access the diagram selection through the routine \texttt{update\_flags}.
The three selection criteria are stored in a derived data type
\texttt{virt\_flags} which has fields \texttt{eval\_0}, \ldots,
\texttt{eval\_2}, in general ranging from zero to the length of the
list given in \texttt{filter.nlo}.
\begin{lstlisting}
use groups
type(virt_flags) :: flags
flags%eval_0=.true. !first group only
flags%eval_1=.false.
flags%eval_2=.false.
call update_flags(flags)
\end{lstlisting}

\paragraph{Additional Extensions}
Some of \GOLEM{}'s functionality is available through the \texttt{extensions}
variable. On top of the already presented options for selecting
a regularisation scheme (by adding the option \texttt{dred}) or
for activating interfaces to
several different reduction libraries (\texttt{samurai}, \texttt{golem95},
\texttt{pjfry}) the user can also add the following options:
\begin{description}
\item[\tt fr5] adds and activates the relevant code for the computation
   of the finite renormalisation of $\gamma_5$ required in the
   \tHV{} scheme as described in Eq.~\eqref{eq:renorm:fr5}.
\item[\tt powhegbox] generates routines for the computation of the
   color and spin correlated Born matrix elements as required by
   \POWHEG{}~\cite{Alioli:2010xd}.
\item[\tt autotools] uses make files which use Autoconf and Automake
   for compilation of the matrix element.
\item[\tt gaugecheck] replaces the polarisation vectors of external
   vector fields by
   \begin{equation}
   \epsilon^\mu(k_i)\to\epsilon^\mu(k_i)+z_ik_i^\mu
   \end{equation}
   where the variable $z_i$ defaults to zero and is accessible in
   the \FORTRAN{} code through the symbol {\tt gauge\it i\tt z}.
\end{description}

%% file: api.tex
The matrix element code generated by \GOLEM{} provides several routines
to transparently access partial or full results of
the amplitude calculation.
Here, we only present a minimal set of routines which can be used to obtain
the set of coefficients $[a_0, c_0, c_{-1}, c_{-2}]$ for a given scale and
a given set of external momenta. The routines, which can be accessed through
the modules \texttt{matrix}\footnote{If a process name was given all
modules are prefixed by the name, e.g. if \texttt{process\_name=pr01}, the
module \texttt{matrix} would be renamed into \texttt{pr01\_matrix}.}
are defined as follows:
\begin{description}
\item[\tt initgolem] This subroutine must be called once before the first
   matrix element evaluation. It initializes all dependent model parameters
   and calls the initialisation routines of the reduction libraries.
   \begin{lstlisting}[language=Fortran]
interface
  subroutine initgolem(init_libs)
  use config, only: ki
  logical, optional, &
 &         intent(in) :: init_libs
  end subroutine
end interface
   \end{lstlisting}
   The optional argument \texttt{init\_libs} can usually be omitted.
   It should be used only when several initialisation calls become necessary,
   but the reduction libraries and loop libraries should be initialized only
   once. All model parameters are accessible as global variables in
   the module \texttt{model} and should be modified (if at all) before
   calling \texttt{initgolem}.
\item[\tt samplitude] This subroutine starts the actual calculation of the
   amplitude for a given phase space point.
   \begin{lstlisting}[language=Fortran]
interface
  subroutine samplitude &
 &        (vecs,scale2,amp,ok,h)
  use config, only: ki
  use kinematics, only: num_legs
  real(ki), dimension(num_legs,4),&
 &         intent(in) :: vecs
  double precision, &
 &         intent(in) :: scale2
  double precision, &
 &         intent(out) :: amp
  logical, optional, &
 &         intent(out) :: ok
  integer, optional, &
 &         intent(in) :: h
  end subroutine
end interface
   \end{lstlisting}
   The first mandatory arguments of this routine are the external momenta
   \texttt{vecs}, where \texttt{vecs(i,:)} contains the momentum of
   the $\mathtt{i}$-th particle as a vector $[E_i, p_i^x, p_i^y, p_i^z]$, 
   and we use in-out kinematics, i.e. $p_1+p_2\to p_3+\ldots +p_N$.
   Maximal numerical stability is achieved if the beam axis is chosen
   along the $z$-axis.
   The second argument, $\mathtt{scale2}=\mu_R^2$,
   is the square of the renormalisation scale.
   As a third argument the routine expects a vector which accepts the
   result in the format $[a_0, c_0, c_{-1}, c_{-2}]$ with the coefficents
   being defined in Eqs. \eqref{eq:amp0:def} and~\eqref{eq:amp1:def}.
   The optional argument \texttt{ok} may be used in order to report
   the outcome of the reconstruction tests in samurai if no rescue method
   has been chosen (see Section~\ref{ssec:reduction-strategies}).
   The last argument allows one to select a single helicity subamplitude;
   the index \texttt{h} runs from zero to the number of helicities minus one.
   The labeling of the helicities is documented for each process
   in the file~\texttt{doc/process.ps}.
\item[\tt exitgolem] This routine should be called once after the last
   amplitude evaluation in the program. It closes all open log files and
   gracefully terminates the reduction and loop libraries.
   \begin{lstlisting}[language=Fortran]
interface
  subroutine exitgolem(exit_libs)
  use config, only: ki
  logical, optional, 
 &         intent(in) :: exit_libs
  end subroutine
end interface
   \end{lstlisting}
   The optional argument \texttt{exit\_libs} should only be set if
   multiple calls to this routine (e.g. for different matrix elements)
   are necessary and the dependent libraries should be terminated only once.
\end{description}

\noindent
A small program which computes the amplitude for a set of phase space points
is automatically generated with the amplitude code in the file
\texttt{test.f90} in the subdirectory \texttt{matrix}.
The script \texttt{config.sh} in the process
directory returns suitable compilation and linking options for the generated
matrix element code.

%% file: blha.tex
\begin{figure}
\begin{center}
\includegraphics[width=\columnwidth]{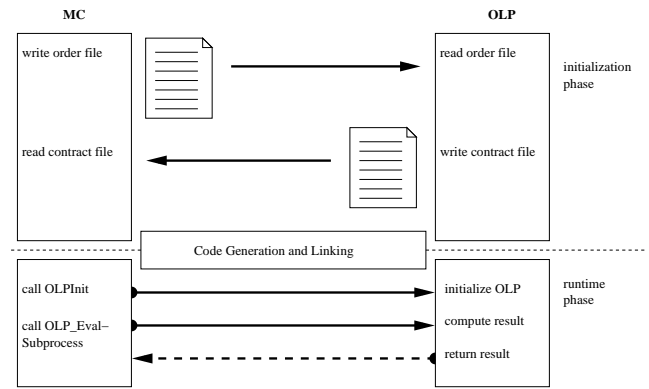}
\end{center}
\caption{Schematic overview over the interaction between Monte Carlo tool
and one-loop program in the Binoth Les Houches Accord.\label{fig:blha}}
\end{figure}

The so-called \emph{Binoth Les Houches Accord} (BLHA)~\cite{Binoth:2010xt}
defines an interface for a standardized communication between
one-loop programs (OLP) and Monte Carlo (MC) tools. The communication
between the two sides is split into two main phases: an initialisation
phase and a runtime phase. During initialisation the two programs
establish an agreement by exchanging a set of files and typically initiate
the code generation. The OLP runtime code is then linked to the MC program
and, during the runtime phase, called through a well-defined set of routines
providing NLO results for the phase space points generated by the MC.
According to this standard, it is the responsibility of the MC program
to provide results for the Born matrix element, for the real emission
and for a suitable set of infrared subtraction terms.
A schematic overview on this procedure is shown in Fig.~\ref{fig:blha}.

\GOLEM{} can act as an OLP in the framework of the BLHA. In the simplest
case, the MC writes an order file --- in this example it is called
\texttt{olp\_order.lh} --- and invokes the script \texttt{gosam.py}
as follows:
\begin{lstlisting}
gosam.py --olp olp_order.lh
\end{lstlisting}
Further, \GOLEM{} specific options can be passed either in a file
or directly at the command line. One can, for example, use autotools
for the compilation by modifying the above line as follows.
\begin{lstlisting}
gosam.py --olp olp_order.lh \
       extensions=autotools
\end{lstlisting}

The contract file is given the extension \texttt{.olc} by default
and would be \texttt{olp\_order.olc} in this example. Alternatively,
the name can be altered using the \texttt{-o} option.

If successful,
the invocation of \texttt{gosam.py} generates a set of files which
can be compiled as before with a generated make file. The BLHA routines
are defined in the \FORTRAN{} module \texttt{olp\_module} but can also
be accessed from \texttt{C} programs\footnote{A header file is provided
in \texttt{olp.h}.}. The routines \texttt{OLP\_Start} and
\texttt{OLP\_EvalSubProcess} are defined exactly as in the BLHA
proposal~\cite{Binoth:2010xt}. For convenience, we extended the interface
by the functions \texttt{OLP\_Finalize()}, which terminates all reduction
libraries, and \texttt{OLP\_Op\-tion(\hspace{0pt}char*,\hspace{0pt}int*)},
which can be used to
pass non-standard options at runtime.
For example, a valid call in \texttt{C} to
adjust the Higgs mass would be
\begin{lstlisting}
int ierr;
OLP_Options("mH=146.78", &ierr);
\end{lstlisting}
A value of one in \texttt{ierr} indicates that the setting was successful.
A value of zero indicates an error.

%% file: model.tex
With a few modifications in the process description files,
\GOLEM{} can immediately make use of model files generated
by either \FEYNRULES{}~\cite{Christensen:2008py}
in the \UFO{} format~\cite{Degrande:2011ua} or by
\LANHEP{}~\cite{Semenov:2010qt}. In both cases, the following
limitations and differences with respect to the default model files,
\texttt{sm} and \texttt{smdiag}, apply:
\begin{itemize}
\item As usual, particles can be specified by their PDG code.
The field names, as used by \QGRAF{}, are \texttt{part$i$} and \texttt{anti$i$}
for the particles with the PDG code $i$ and $-i$ respectively. For example,
the $W^+$ and the $W^-$ boson would be called \texttt{part24} and
\texttt{anti24}.
\item All model parameters are prefixed by the letters \texttt{mdl} in order
to avoid name clashes with existing variable names in the matrix element code.
\item The variable \texttt{model.options} and the extension \texttt{fr5}
are not guaranteed to work with models other than the built-in models.
\end{itemize}

\paragraph{Importing models in the \UFO{} format}
A model description in the \UFO{} format consists of a \PYTHON{} package
stored in a directory. In order to import the model into \GOLEM{} one needs
to set the \texttt{model} variable in the input card to specify the keyword \texttt{FeynRules}
in front of the directory name, where we assume that
the model description is in the directory \texttt{\$HOME/models/MSSM\_UFO}.
\begin{lstlisting}
model=FeynRules,$HOME/models/MSSM_UFO
\end{lstlisting}

\paragraph{Importing models in the \LANHEP{} format}
\LANHEP{} model descriptions consist of a set of plain text files in the
same directory with a common numbering (such as
\texttt{func4}\hspace{0pt}\texttt{.mdl},
\texttt{lgrng4.mdl}, \texttt{prtcls4.mdl},
\texttt{vars4.mdl}). A \LANHEP{} model can be loaded by specifying the
path and the common number in the \texttt{model} variable.
Assuming the files are situated in the
directory \texttt{\$HOME/}\hspace{0pt}\texttt{models/}\hspace{0pt}%
\texttt{MSSM\_LHEP} one would set the variable as follows.
\begin{lstlisting}
model=$HOME/models/MSSM_LHEP,4
\end{lstlisting}
Details about the allowed names for the table columns are described in
the \GOLEM{} reference manual. 
Precompiled \texttt{MSSM\_UFO} and \texttt{MSSM\_LHEP} files can also 
be found in the subdirectory \texttt{examples/model}.

%% file: wjet.tex
In Section~\ref{ssec:blha} the BLHA interface of \GOLEM{} was presented. This interface allows one to link the program to a Monte Carlo event generator, which is, in general, responsible for supplying the missing ingredients for a complete NLO calculation of a physical cross section. 
Among the different general purpose Monte Carlo event generators, 
\SHERPA{}\cite{Gleisberg:2008ta} is one of those which offers these tools: 
computing the LO cross section, 
the real corrections with both the subtraction terms 
and the corresponding integrated counterparts~\cite{Krauss:2001iv,Gleisberg:2007md,Schonherr:2008av}. 
Furthermore, \SHERPA{} offers the possibility to match a NLO calculation with a parton shower~\cite{Hoche:2010pf,Hoeche:2011fd}. Using the BLHA interface, we linked \GOLEM{} with \SHERPA{} to compute the physical cross section for $W^{-}+1$ jet at NLO.

The first steps to perform this linking is to write a \SHERPA{} input card for the desired process. Instructions and many examples on how to write this can be found in the on-line manual~\cite{Sherpa1.3.1:online}. Running the code for the first time will produce an order file \emph{OLE\_order.lh} which contains all the necessary information for \GOLEM{}, to produce the desired code for the loop part of the process. This includes a list of all partonic subprocesses needed. In parallel to the production of the needed \SHERPA{} libraries with the provided script, one can at this point run the \texttt{gosam.py} command with the flag \texttt{--olp} and the correct path to the order file as explained in Section~\ref{ssec:blha}. Further options may be specified. Among them it is useful to have a second, \GOLEM{}-specific, input card with all the important \GOLEM{} options. Since, at the end, \SHERPA{} needs to be linked to a dynamic library, it is convenient to run \GOLEM{} with the \emph{autotools} extension, which allows the direct creation of both static and dynamic libraries, together with the test routine \texttt{test}. The \texttt{gosam.py} script creates all the files needed for interfacing \GOSAM{} with the Monte Carlo event generator together with the code for the one-loop computation of all needed subprocesses, and a makefile to run them. The different parton-level subprocesses are contained in different subdirectories. At this point the user simply has to run the makefile to generate and compile the code. Once the one-loop part of the code is ready, the produced shared library must be added to the list of needed libraries in the \SHERPA{} input card as follows.
\begin{lstlisting}
SHERPA_LDADD = LHOLE golem_olp;
\end{lstlisting}
With this operation the generation of the code is completed. The evaluation of the process and the physical analysis can then be performed at the 
user's discretion following the advice given in the \SHERPA{} on-line documentation~\cite{Sherpa1.3.1:online}.

\begin{figure*}[hbt]
\centering
\subfigure[Inclusive transverse momentum of the leading jet for $W^{-}+1$~jet production at LHC with $\sqrt{s}=7$~TeV.]{\label{fig:gosamsherpa:pt}%
\includegraphics[width=0.45\textwidth]{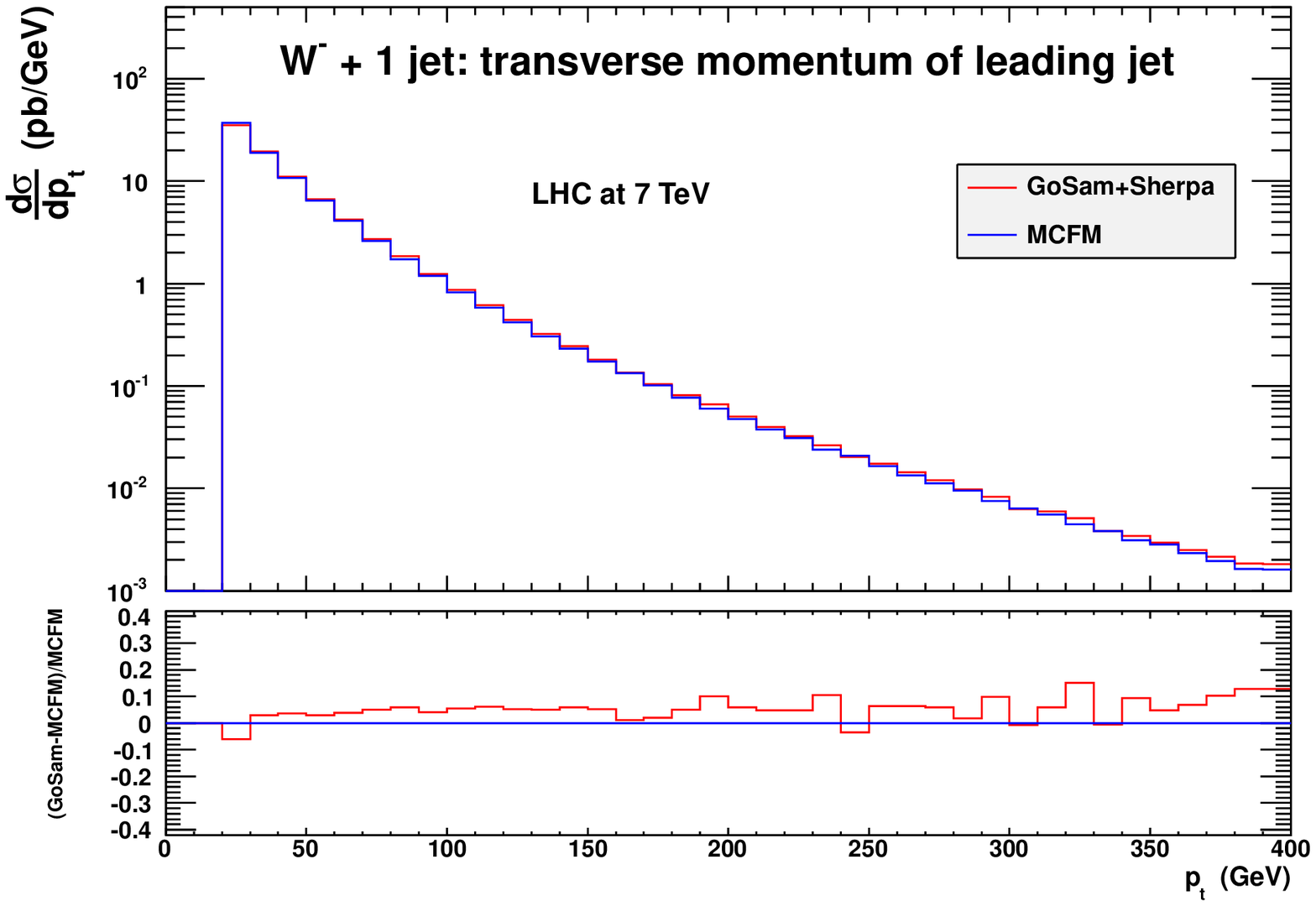}}
\subfigure[Inclusive pseudorapidity of the leading jet for $W^{-}+1$~jet production
at LHC with $\sqrt{s}=7$~TeV.]{\label{fig:gosamsherpa:eta}%
\includegraphics[width=0.45\textwidth]{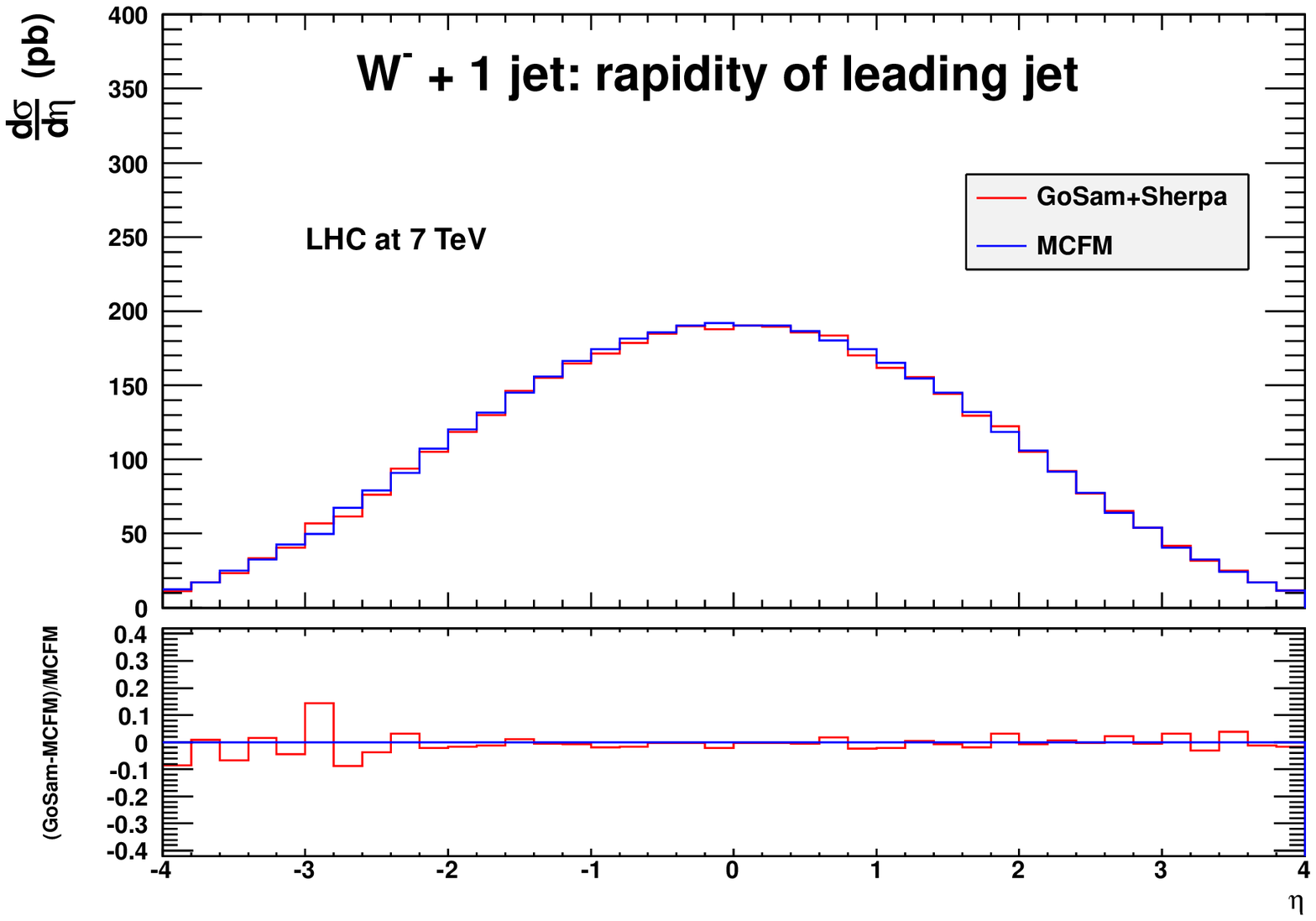}}
\caption{NLO calculation of $W^{-}+1$ jet production at LHC using \GOLEM{} 
interfaced with \SHERPA{} via the BHLA interface. The comparison to MCFM is also shown.}
\end{figure*}

We tested the BLHA interface by computing $W^{-}+1$ jet and producing distributions for several typical observables. 
In Figs.~\ref{fig:gosamsherpa:pt} and~\ref{fig:gosamsherpa:eta} the inclusive transverse momentum and rapidity of the jets is shown. 
These distributions were compared with similar ones produced using the  program
 \MCFM{}~\cite{Campbell:1999ah,Campbell:2011bn}, and perfect agreement was found.


%% file: wjetew.tex
As a first example of an electroweak calculation, we computed the virtual one-loop corrections to  
$u \bar{d} \to W g$. 
A complete analytical calculation for this process was presented in Ref.~\cite{Kuhn:2007cv}.

\begin{table*}
\begin{center}
\begin{kinematics}
$u$ & 500 &  0 & 0 & 500 \\
$\bar{d}$ & 500 &  0 & 0 & 500 \\
$W$ & 503.23360778049988 & 110.20691318538486 & 441.95397288433196 & -198.26237811718670 \\
$g$ & 496.76639221950012 & -110.20691318538488 & -441.95397288433202 & 198.26237811718664 
\end{kinematics}
\end{center}
\caption{Kinematic point used in $pp\to W^\pm+j$, EW.}\label{tab:kin:Wjetew}
\end{table*}

\begin{center}
\begin{parameters}
$M_Z$ & 91.1876 & $M_W$ & 80.419 \\
$\cos\theta_w$ & 0.88156596117995394232 & $\mu$ & $M_W$
\end{parameters}
\end{center}

\noindent For the kinematic point given in Tab.~\ref{tab:kin:Wjetew}
and the above parameters we obtain the following result:
\begin{center}
\small
\begin{tabular}{|lll|}
\hline
\multicolumn{3}{|c|}{{result $u \bar{d} \to W g$}}\\
\hline
\multicolumn{2}{|l}{{$a_0$}} & 2.812364835883295 \\
\multicolumn{2}{|l}{{$c_0/a_0$ unren.}} & -94.52525523327047 \\ 
\multicolumn{2}{|l}{{$c_{-1}/a_0$ unren.}} & 17.84240236996827 \\ 
\multicolumn{2}{|l}{{$c_{-2}/a_0$ unren.}} &-0.5555555555555560 \\
\hline
\multicolumn{3}{|c|}{{renormalized}}\\
\hline
&\GOLEM &Eqs.(67,70) of Ref.~\cite{Kuhn:2007cv} \\
\hline
$c_{-1}/a_0$ &  4.743825167813529 &  4.7438251678146885 \\
$c_{-2}/a_0$ &-0.5555555555555560 & -0.5555555555555555 \\ 
\hline
\end{tabular}
\end{center}

The poles have been renormalized using Eqs.(49)-(64) in Sections~3.3 and~3.4 of~\cite{Kuhn:2007cv}. 
Our result is agreement with Eqs.(67),(70) of Ref.~\cite{Kuhn:2007cv} and with Ref.~\cite{Gehrmann:2010ry} for the infrared divergences that remain after renormalisation.

%% file: yyyy.tex
The process $\gamma\gamma\to\gamma\gamma$ in the Standard Model first arises
at the one-loop order, and proceeds through a closed loop of fermions
and $W$ bosons.
Of the 16 helicity amplitudes contributing to it, only three are
independent and their analytic expressions can be found in
\cite{Gounaris:1999gh}.
The pure QED contribution, involving a fermion loop, is contained in
{\tt samurai-1.0}
\cite{Mastrolia:2010nb} and will not be repeated here. Instead,
we show the results of the $W$-loop contribution to the independent
helicity amplitudes, as an example of EW corrections that can be
handled with \GOSAM{}.

\begin{table*}[hbtp]
\begin{center}
\begin{kinematics}
$\gamma$& 500&  0&  0&  500\\
$\gamma$& 500&  0&  0& -500\\
$\gamma$& 500& 436.6186300198938284 &   -59.1784256571505765  &
236.3516148798047425 \\
$\gamma$& 500&-436.6186300198938284 &   59.1784256571505765  &
-236.3516148798047425
\end{kinematics}
\end{center}
\caption{Kinematic point used in $\gamma\gamma\to\gamma\gamma$.}
\label{tab:kin:yyyy}
\end{table*}

\begin{center}
\begin{parameters}
$\sqrt{s}$ & 1000  & $\mu$ & $\sqrt{s}$ \\
$M_W$ & 80.376  & $e$ & 1
\end{parameters}
\end{center}
With the above parameters and the kinematics of Tab.~\ref{tab:kin:yyyy}
we obtain the following results.
\begin{center}
\small
\begin{tabular}{|lll|}
\hline
\multicolumn{3}{|c|}{result $\gamma\gamma\to\gamma\gamma$ (EW)}\\
\hline
&\GOLEM (dred)&Refs.\cite{Gounaris:1999gh}\\
\hline
$ |{\cal M}_{++++}|$ & 12.02541904626610 & 12.025419045962 \\
$ |{\cal M}_{++-+}|$ & 7.380406043429961 & 7.3804060437434\\
$ |{\cal M}_{++--}|$ & 982.7804939723322 & 982.78049397093 \\
\hline
\end{tabular}
\end{center}

%% file: neutralino.tex
As an example for the usage of {\tt GoSam} with a model file different from the 
Standard Model we calculated the QCD corrections to neutralino pair production
in the MSSM. The model file has been imported via the interface {\tt UFO} 
(Universal FeynRules Output) \cite{Degrande:2011ua} which facilitates  
the import of Feynman rules generated by {\tt FeynRules} \cite{Christensen:2008py}
to programs generating one-loop amplitudes. 
To import such files within the {\tt GoSam} setup, all the user has to do 
is to give the path to the corresponding model file in the input card.

For this example, we combined the one-loop amplitude with the 
real radiation corrections to obtain results for differential cross sections.
A calculation of neutralino pair production for the LHC presenting total cross sections
at NLO is given in \cite{Beenakker:1999xh}. 

For the infrared subtraction terms the program {\tt Mad\-Di\-pole}
\cite{Frederix:2008hu,Frederix:2010cj} is used,  
the real emission part is calculated using Mad\-Graph/Mad\-Event \cite{Alwall:2007st}.
The virtual matrix element is renormalized in the $\overline{MS}$ scheme, 
while massive particles are treated in the on-shell scheme. 
The renormalisation terms specific to the massive MSSM particles have been added manually.

\begin{figure}[t]
\centering
\subfigure{\includegraphics[width=\columnwidth]{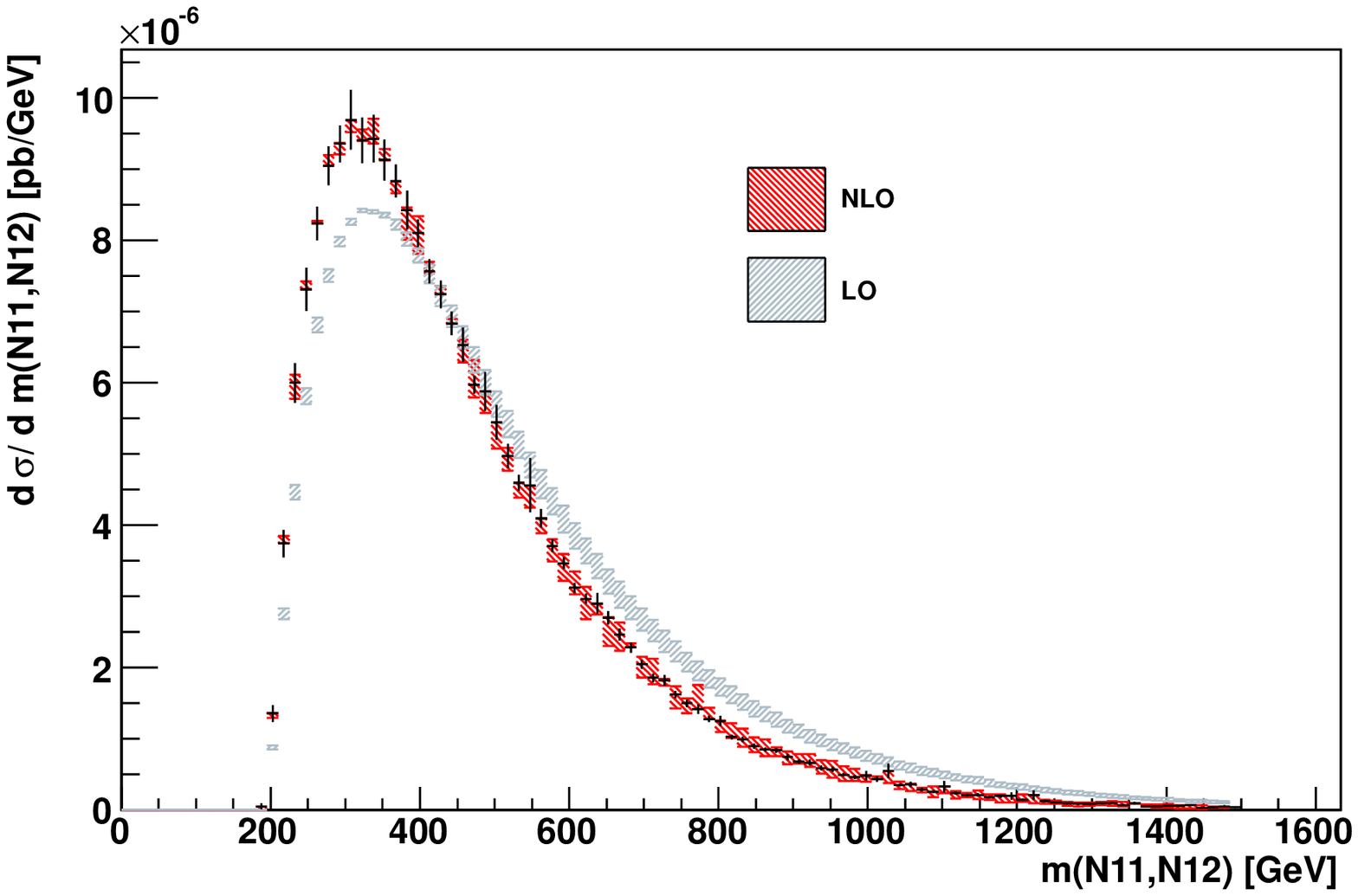}}
\caption{Comparison of the NLO and LO $m_{\chi_{1}^{0} \chi_{1}^{0}}$ distributions for the process $pp \rightarrow \chi_{1}^{0} \chi_{1}^{0}$
with a jet veto on jets with $p^{jet}_{T} > 20$ GeV and $\eta<4.5$.
The band gives the dependence of the result on $\mu = \mu_{F} = \mu_{R}$ between
$\mu_{0}/2$ and $2 \mu_{0}$. We choose $\mu_{0} = M_{Z}$. The black line gives the bin
error for the value at the central scale.}
\label{fig:qqNNm12JVfull}
\end{figure}

In Fig.\,\ref{fig:qqNNm12JVfull} we show the differential cross section for the 
$m_{\chi_{1}^{0} \chi_{1}^{0} }$ invariant  mass,
where we employed a jet veto to suppress large contributions from 
the channel $qg \rightarrow \chi_{1}^{0} \chi_{1}^{0} q$ which opens up 
at order $\alpha^2 \alpha_s$, but for large $p_{T}^{jet}$ belongs to
the distinct process of neutralino pair
plus one hard jet production at leading order.
We used $N_f=5$ massless quark flavours and the MSTW08 \cite{Martin:2009iq} parton distribution functions.
For the SUSY parameters we use the modified benchmarks point SPS1amod suggested in
\cite{Feigl:2011sw}, and we use $\sqrt{s}=7$\,TeV.

For reference, we also give the result for the unrenormalised 
amplitude at one specific phase space point for 
$u\bar{u} \to \chi_{1}^{0} \chi_{1}^{0}$ in the DRED scheme, 
using the following parameters and momenta:

\begin{table*}
\begin{center}
\begin{kinematics}
$u$ &1000 & 0& 0& 1000 \\
$\bar{u}$ &1000 &0 &0 & -1000 \\
$\chi_{1}^{0}$ & 1000&42.3752677206678996 &115.0009952646289548 & 987.7401101322898285 \\
$\chi_{1}^{0}$ &1000 &-42.3752677206678996 &-115.0009952646289548 &-987.7401101322898285 \\
\end{kinematics}
\end{center}
\caption{Kinematic point used in $pp\to\chi_1^0\chi_1^0$ in the
MSSM.}\label{tab:kin:xixi}
\end{table*}

\begin{center}
\begin{parameters}
$M_Z$ & 91.1876 & $\Gamma_Z$ & 0\\
$M_W$ & 79.829013 & $\sin^2\theta_w$ & $1-M_W^2/M_Z^2$  \\
$\mu$ & $M_Z$ & $N_f$ & 1 \\
$g_s$ & 1 &$\alpha_w^{-1}$ &  127.934\\
$M_{\chi_{1}^{0}}$& 96.6880686 & $M_{\tilde{g}}$ & 607.713704 \\
$M_{\tilde u_L}$ & 561.119014 &$M_{\tilde u_R}$ & 549.259265\\
$M_{h_0}$ & 110.899057 & $M_{H_0}$ & 399.960116\\
\end{parameters}
\end{center}

All widths have been set to zero; for further settings we refer to the 
model parameter files contained in the subdirectory {\tt examples/model/MSSM\_UFO}.
We have checked that the pole terms of       
the renormalised amplitude cancel with the infrared poles from Mad\-Di\-pole.
For the phase space point given in Tab.~\ref{tab:kin:xixi} we obtain
the following numbers.
\begin{center}
\small
\begin{tabular}{|ll|}
\hline 
\multicolumn{2}{|c|}{\small \GOLEM{} result $u\bar{u}\to \chi_{1}^{0} \chi_{1}^{0}$}\\
\hline
$a_{0}$ & 0.8680577964243597$\cdot 10^{-3}$ \\
$c_0/a_0$ & -31.9136615197871   \\
$c_{-1}/a_0$ & 13.4374663711899  \\
$c_{-2}/a_0$ & 2.6666666666667    \\
\hline
\end{tabular}

\end{center}


%% file: eeey.tex
As an example of a QED calculation, we compared the virtual QED corrections for the process $e^+e^-\to e^+e^-\gamma$ with the results provided in~\cite{Actis:2009uq}.
The results compared in the table are the bare unrenormalised amplitudes in the 't~Hooft Veltman scheme. No counterterms or subtraction terms have been added to the result.

\begin{table*}[ptb]
\begin{center}
\begin{kinematics}
$e^+$ (in) & 0.5 & 0 &0.4999997388800458 & 0\\
$e^-$ (in) &0.5 & 0 &-0.4999997388800458 & 0\\
$e^+$ (out) & 0.1780937847558600& −0.1279164180985903& 0.05006809884093004 & −0.1133477415216646\\
$e^-$ (out)& 0.3563944406457374& −0.02860530642319879 &−0.1832142729949070 &−0.3043534176228102 \\
$\gamma$  & 0.4655117745984024& 0.1565217245217891& 0.1331461741539769 &  0.4177011591444748
\end{kinematics}
\end{center}
\caption{Kinematic point used in $e^+e^-\to e^+e^-\gamma$.}
\label{tab:kin:eeeey}
\end{table*}

\begin{center}
\begin{parameters}
$\sqrt{s}$ & 1.0 & $\alpha$ & $7.2973525376\cdot10^{-3}$ \\
$\mu$ & $\sqrt{s}$ & $m_e$ & $0.51099891\cdot10^{-3}$ \\ 
\end{parameters}
\end{center}

Using the parameters given above and the kinematics of
Tab.~\ref{tab:kin:eeeey} we obtain the following results.
\begin{center}
\small
\begin{tabular}{|lll|}
\hline
\multicolumn{3}{|c|}{result $e^+e^-\to e^+e^-\gamma$}\\
\hline
& \GOLEM{} & Ref.~\cite{Actis:2009uq} \\
\hline
$a_0$ &  $0.7586101468103622$ & $  0.7586101468103619 $\\
$c_0/a_0$    & $  0.5005827938274887 $ & $  0.5005828268263969 $ \\
$c_{-1}/a_0$ & $  0.0474506407008029  $ & $ 0.0474506427003504 $ \\
$c_{-2}/a_0$ & $0$ & $0$ \\
\hline 
\end{tabular}
\end{center}

%% file: ttH.tex
This process has been compared with the results given
in~\cite{Hirschi:2011pa}.
The partonic subprocesses $u\bar{u}\rightarrow t\bar{t}H$ and $gg\rightarrow t\bar{t}H$ where
computed both in the \tHV{} scheme and in
dimensional reduction and the fully renormalised results were successfully compared as an 
internal consistency check. 
Apart from wave function renormalisation and mass counterterms, 
Yukawa coupling renormalisation is also needed here. 
Yukawa coupling counterterms are in this case equal to the wave function counterterms. 
The Yukawa top mass is set equal to its pole mass.

\begin{center}
\begin{parameters}
$\sqrt{s}$ & 500.0 & $N_f$ & 5 \\
$\mu$ & $m_t$ & $N_{f,h}$ & 1 \\
$m_t$ & 172.6
& $\alpha_s$ & 0.1076395107858145 \\
$m_H$ & 130 &
$v$ & 246.21835258713082
\end{parameters}
\end{center}

The kinematics used to obtain the results below is given
in Tab.~\ref{tab:kin:ttH}.
The results are given in the \tHV{} scheme,
and are fully renormalised.
\begin{center}
\small
\begin{tabular}{|lll|}
\hline
\multicolumn{3}{|c|}{result $u\bar{u}\to t\bar{t} H$}\\
\hline
& \GOLEM{} & Ref.~\cite{Hirschi:2011pa} \\
\hline
$a_0\cdot10^4$ &  $2.200490364806190$ & $2.2004904613782828$\\
$c_0/a_0$    & $-15.29615178164782$ & $-15.29615211731521$ \\
$c_{-1}/a_0$ & $-1.640361500121837$ & $-1.640361536072381$ \\
$c_{-2}/a_0$ & $-2.666666666666666$ & $-2.666666725182165$ \\
\hline 
\end{tabular}

\bigskip

\begin{tabular}{|lll|}
\hline
\multicolumn{3}{|c|}{result $gg\to t\bar{t} H$}\\
\hline
& \GOLEM{} & Ref.~\cite{Hirschi:2011pa} \\
\hline
$a_0\cdot10^5$ &  $6.127399805961155$ & $6.127400074872043$\\
$c_0/a_0$    & $9.006680638719660$  & $9.006680836410272$ \\
$c_{-1}/a_0$ & $2.986347664537282$  & $2.9863477301662056$ \\
$c_{-2}/a_0$ & $-6.000000000000004$ & $-6.000000131659877$ \\
\hline
\end{tabular}
\end{center}

\begin{table*}[ptb]
\begin{center}
\begin{kinematics}
$u/g$ & 250.0 & 0.0 & 0.0 & 250.0 \\
$\bar{u}/g$ & 250.0 & 0.0 & 0.0 &-250.0 \\
$H$ & 136.35582793693018 & 15.133871809486299 &  27.986733991031045 & 26.088703626953386 \\
$t$ & 181.47665951104506 & 20.889486679044587 & -50.105625289561424 & 14.002628607367491 \\
$\bar{t}$ & 182.16751255202476 & -36.023358488530903 & 22.118891298530357 & -40.091332234320859 \\
\end{kinematics}
\end{center}
\caption{Kinematic point used in $pp\to t\bar{t}H$.}
\label{tab:kin:ttH}
\end{table*}

On an Intel Core i7 950 at 3\,GHz the evaluation of a single phase space
point took~44\,ms in the $u\bar{u}$ channel and~223\,ms in the $gg$ channel.
The code was compiled with \texttt{gfortran} without optimisations.

%% file: ttZ.tex
This amplitude, fully renormalised, has been compared with the results given
in~\cite{Kardos:2011na}.

\begin{table*}[ptb]
\begin{center}
\begin{kinematics}
$g$ & 7000.0 & 0.0 & 0.0 & 7000.0 \\
$g$ & 7000.0 & 0.0 & 0.0 &-7000.0 \\
$t$ & 6270.1855170414337 & -4977.7694025303863 & 806.93726196887712   &     3725.2619580634337 \\
$\bar{t}$ & 6925.5258180925930 &  5306.3374282745517 & -1281.8763412410237   &   -4258.3185872039012   \\
$Z$ & 804.28866486597315 & -328.56802574416463 & 474.93907927214622   & 533.05662914046729  \\
\end{kinematics}
\end{center}
\caption{Kinematic point used for $gg\to t\bar{t}Z$.}
\label{tab:kin:ttZ}
\end{table*}
\begin{center}
\begin{parameters}
$g_s$ & 1 & $G_F$ &  0.0000116639 \\
$\mu$ & $m_t$ & $N_f$ & 5 \\
$m_t$ & 170.9 & $M_W$ & 80.45 \\
$M_Z$ & 91.18 &  \\
\end{parameters}
\end{center}

\noindent The kinematics used to obtain the results below is
given in Tab.~\ref{tab:kin:ttZ}.
\begin{center}
\small
\begin{tabular}{|lll|}
\hline
\multicolumn{3}{|c|}{result $gg\to t\bar{t} Z$}\\
\hline
& \GOLEM{} & Ref.~\cite{Kardos:2011na}\\
\hline
$a_0\cdot 10^6$       & $ 0.1531395190212139$ & $ 0.1531395190212831$\\
$c_0/a_0$   & $ -204.9208290898557 $  & $ -204.920829867328$ \\
$c_{-1}/a_0$ & $ 50.62939646427283$  & $ 50.6293965717156$ \\
$c_{-2}/a_0$ & $ -5.999999999999997$ & $ -6.00000000000003$ \\
\hline
\end{tabular}
\end{center}

\noindent The evaluation of a single phase space
point took~$1433$\, ms on a 2\,GHz processor.
The code was compiled with \texttt{gfortran -O2}.

%% file: bbbb.tex
A detailed discussion of this process can be found in \cite{Greiner:2010ci,Binoth:2010pb}.
In this section we focus on the parts that are relevant in the context of the virtual corrections. In particular we compared
our result to the one given in \cite{vanHameren:2009dr}, which is the fully renormalised amplitude including
the mass counterterms for the top-quark contribution.

\begin{table*}[ptb]
\begin{center}
\begin{kinematics}
$u/g$ & 250.0 & 0.0 & 0.0 & 250.0 \\
$\bar{u}/g$ & 250.0 & 0.0 & 0.0 & -250.0 \\
$b$ &  147.5321146846735 & 24.97040523056789 
    & -18.43157602837212 & 144.2306511496888 \\
$\bar{b}$ &   108.7035966213640 & 103.2557390255471 &
            -0.5484684659584054 & 33.97680766420219 \\
$b$ &  194.0630765341365 & -79.89596300367462
    &  7.485866671764871 & -176.6948628845280 \\
$\bar{b}$ & 49.70121215982584 & -48.33018125244035
          & 11.49417782256567 & -1.512595929362970
\end{kinematics}
\end{center}
\caption{Kinematic point used in $pp\to b\bar{b}b\bar{b}$.}
\label{tab:kin:bbbb}
\end{table*}

\begin{center}
\begin{parameters}
$\sqrt{s}$ & 500 & $N_f$ & 5 \\
$\mu$ & $\sqrt{s}$ & $N_{f,h}$ & 1 \\
$m_t$ & 174 & $m_b$ & 0 \\
$\Gamma_t$ & 0 & $g_s$ & 1 
\end{parameters}
\end{center}
The results below are obtained for the phase space
point of Tab.~\ref{tab:kin:bbbb} using the above parameters.
\begin{center}
\small
\begin{tabular}{|lrr|}
\hline
\multicolumn{3}{|c|}{\small result $gg \to b \bar{b} b \bar{b}$}\\
\hline
&\multicolumn{1}{l}{\GOLEM}
&\multicolumn{1}{l|}{Ref.~\cite{vanHameren:2009dr}}\\
\hline
$a_0\cdot10^6$ & $1.022839601391936$ & $1.022839601391910$\\
$c_0/a_0$    & $-36.97653243659754$ & $-36.97653243473214$\\
$c_{-1}/a_0$ & $-34.01491655155776$ & $-34.01491655142099$ \\
$c_{-2}/a_0$ & $-11.33333333333512$ & $-11.33333333333343$ \\
\hline 
\end{tabular}

\bigskip

\begin{tabular}{|lrr|}
\hline
\multicolumn{3}{|c|}{\small result $u\bar{u} \to b \bar{b} b \bar{b}$}\\
\hline
&\multicolumn{1}{l}{\GOLEM}
&\multicolumn{1}{l|}{Ref.~\cite{vanHameren:2009dr}}\\
\hline
$a_0\cdot10^9$ & $5.753293428094349$ & $5.753293428094391$\\
$c_{0}/a_0$  & $-22.19223384585620$ & $-22.19223384564902$ \\
$c_{-1}/a_0$ & $-20.89828996870689$ & $-20.89828996857439$ \\
$c_{-2}/a_0$ & $-8.000000000000199$ & $-8.000000000000037$ \\
\hline
\end{tabular}
\end{center}

On an Intel Xeon E7340 the running time for the calculation of a
single phase space point was $19.6$\;s for the gluon
initiated channel and $440$\;ms for the quark channel.

%% file: ttbb.tex
This process has been compared with the results given
in~\cite{vanHameren:2009dr}.
We have set up the process both in the \tHV{} scheme and in
dimensional reduction and successfully compared the fully renormalised
results as an internal consistency check. The results below are given
in the \tHV{} scheme, and only the counterterms for
$\vert\mathcal{M}\vert_{\text{ct, $\delta m_t$}}^2$ are included.

%
%
%

\begin{table*}[ptb]
\begin{center}
\begin{kinematics}
$u/g$ & 250.0 & 0.0 & 0.0 & 250.0 \\
$\bar{u}/g$ & 250.0 & 0.0 & 0.0 &-250.0 \\
$t$ & 190.1845561691092 & 12.99421901255723  &
       -9.591511769543683 & 75.05543670827210 \\
$\bar{t}$ & 182.9642163285034 & 53.73271578143694  &
      -0.2854146459513714 & 17.68101382654795 \\
$b$ & 100.9874727883170 &-41.57664370692741  &
        3.895531135098977 &-91.94931862397770 \\
$\bar{b}$ & 25.86375471407044 & -25.15029108706678  &
           5.981395280396083 & -0.7871319108423604
\end{kinematics}
\end{center}
\caption{Kinematic point used in $pp\to t\bar{t}b\bar{b}$.}
\label{tab:kin:ttbb}
\end{table*}

\begin{center}
\begin{parameters}
$\sqrt{s}$ & 500.0 & $N_f$ & 5 \\
$\mu$ & $\sqrt{s}$ & $N_{f,h}$ & 1 \\
$m_t$ & 174.0 & $m_b$ & 0.0 \\
$\Gamma_t$ & 0.0 & $g_s$ & 1.0 
\end{parameters}
\end{center}

Using the above parameters and the phase space point
of Tab.~\ref{tab:kin:ttbb} we obtain the following results.

\begin{center}
\small
\begin{tabular}{|lrr|}
\hline
\multicolumn{3}{|c|}{\small result $u\bar{u}\to t\bar{t} b\bar{b}$}\\
\hline
& \multicolumn{1}{l}{\GOLEM{}}
& \multicolumn{1}{l|}{Ref.~\cite{vanHameren:2009dr} }\\
\hline
$a_0\cdot10^8$ &  $2.201164677187755$ & $2.201164677187727$\\
$c_0/a_0$    & $8.880263116574282$ & $8.880263117410131$ \\
$c_{-1}/a_0$ & $-4.730495922109534$ & $-4.730495921691266$ \\
$c_{-2}/a_0$ & $-5.333333333333468$ & $-5.333333333333190$ \\
\hline
\end{tabular}

\bigskip

\begin{tabular}{|lrr|}
\hline 
\multicolumn{3}{|c|}{\small result $gg\to t\bar{t} b\bar{b}$}\\
\hline
& \multicolumn{1}{l}{\GOLEM{}}
& \multicolumn{1}{l|}{Ref.~\cite{vanHameren:2009dr} }\\
\hline
$a_0\cdot10^8$ &  $8.279470201927135$ & $8.279470201927128$\\
$c_0/a_0$    & $21.83922035777929$ & $21.83922035648926$ \\
$c_{-1}/a_0$ & $-12.59181277770347$ & $-12.59181277853837$ \\
$c_{-2}/a_0$ & $-8.666666666666764$ & $-8.666666666666549$ \\
\hline
\end{tabular}
\end{center}
\bigskip

On an Intel Core i7 950 at 3\,GHz the evaluation of a single phase space
point took~393\,ms in the $u\bar{u}$ channel and 12{.}3\,s in the $gg$ channel.
The code was compiled with \texttt{gfortran} without optimisations.

%% file: wwbb.tex
The subprocesses
$u\bar{u}\to W^+W^-b\bar{b}$ and $gg\to W^+W^-b\bar{b}$ have been 
calculated both in
\cite{vanHameren:2009dr} and~\cite{Hirschi:2011pa}. Accordingly, the 
results below are given
in the \tHV{} scheme, where only the counterterms for
$\vert\mathcal{M}\vert_{\text{ct, $\delta m_t$}}^2$ are included.

\begin{table*}[ptb]
\begin{center}
\begin{kinematics}
$u/g$ & 250.0 & 0.0 & 0.0 & 250.0 \\
$\bar{u}/g$ & 250.0 & 0.0 & 0.0 & -250.0 \\
$W^+$ &  154.8819879118765 & 22.40377113462118
      & -16.53704884550758 & 129.4056091248114 \\
$W^-$ &  126.4095336206695 & 92.64238702192333 
      &-0.4920930146078141 & 30.48443210132545 \\
$b$   &  174.1159068988160 &-71.68369328357026
      &  6.716416578342183 &-158.5329205583824 \\
$\bar{b}$
      &  44.59257156863792 &-43.36246487297426
      &  10.31272528177322 &-1.357120667754454
\end{kinematics}
\end{center}
\caption{Kinematic point used in $pp\to W^+W^-b\bar{b}$.}
\label{tab:kin:wwbb}
\end{table*}

\begin{center}
\begin{parameters}
$\sqrt{s}$ & 500.0 & $N_f$ & 5 \\
$\mu$ & $\sqrt{s}$ & $N_{f,h}$ & 1 \\
$m_t$ & 174.0 & $m_b$ & 0 \\
$\Gamma_t$ & 0 & $g_s$ & 1\\
$M_Z$ & 91.188 & $\Gamma_Z$ & 2.44140351 \\
$M_W$ & 80.419 & $\Gamma_W$ & 0 \\
$1/\alpha$ & 132.50686625 &
\end{parameters}
\end{center}

\noindent
With the above parameters and the kinematics defined
in Tab.~\ref{tab:kin:wwbb} we obtain the following results.

\begin{center}
\small
\begin{tabular}{|lrr|}
\hline
\multicolumn{3}{|c|}{result $gg \to W^+W^-b \bar{b}$}\\
\hline
&\multicolumn{1}{l}{\GOLEM}
&\multicolumn{1}{l|}{Ref.~\cite{Hirschi:2011pa}}\\
\hline
$a_0\cdot10^8$ & $1.549796787502985$ & $1.549795815702494$\\
$c_0/a_0$    & $-17.80558461276584$ & $-17.80558440908488$\\
$c_{-1}/a_0$ & $-19.61125131175888$ & $-19.611251301307803$ \\
$c_{-2}/a_0$ & $-8.666666666666668$ & $-8.66666666666661$ \\
\hline 
\end{tabular}

\bigskip

\begin{tabular}{|lrr|}
\hline\multicolumn{3}{|c|}{result $u\bar{u} \to W^+W^- b \bar{b}$}\\
\hline
&\multicolumn{1}{l}{\GOLEM}
&\multicolumn{1}{l|}{Ref.~\cite{Hirschi:2011pa}}\\
\hline
$a_0\cdot10^8$ & $2.338048681706755$ & $2.338048676370483$\\
$c_{0}/a_0$  & $-5.936151367348438$ & $-5.936151368788066$ \\
$c_{-1}/a_0$ & $-10.44868110371249$ & $-10.44868110378090$ \\
$c_{-2}/a_0$ & $-5.333333333333312$ & $-5.333333333333336$ \\
\hline
\end{tabular}
\end{center}

%% file: udwggg.tex
%
%
%
%
%
%
%
The amplitude $u\bar{d}\to W^+ggg$ is an important channel in the calculation
of the process $pp\to W^++3\,\text{jets}$. The QCD corrections to this process
have been presented in Refs.~\cite{Berger:2009ep,Berger:2009zg,KeithEllis:2009bu,Melnikov:2009wh}. 

The subprocess with
one quark pair and three gluons consists of more than 1500 Feynman diagrams.
We have computed the amplitude including the leptonic decay of the $W$-boson
and compared our result to~\cite{vanHameren:2009dr}.

\begin{table*}[ptb]
\begin{center}
\begin{kinematics}
$u$       & 250.0 & 0.0 & 0.0 & 250.0 \\
$\bar{d}$ & 250.0 & 0.0 & 0.0 & -250.0 \\
$W^+$     & 162.5391101447744 & 23.90724239064912
          &-17.64681636854432 & 138.0897548661186 \\
$g$       & 104.0753327455388 & 98.85942812363483
          &-0.5251163702879512 & 32.53017998659339 \\
$g$       & 185.8004692730082 &-76.49423931754684
          & 7.167141557113385 &-169.1717405928078 \\
$g$       & 47.58508783667868 &-46.27243119673712
          & 11.00479118171890 &-1.448194259904179
\end{kinematics}
\end{center}
\caption{Kinematic point used in $u\bar{d}\to W^+ggg$.}
\label{tab:kin:wggg}
\end{table*}

\begin{center}
\begin{parameters}
$\sqrt{s}$ & 500.0 & $N_f$ & 5 \\
$\mu$ & $\sqrt{s}$ & $N_{f,h}$ & 1 \\
$m_t$ & 174.0 & $M_Z$ & 91.188 \\
$\Gamma_t,\Gamma_W,\Gamma_Z$ & 0.0 & $M_W$ & 80.419 \\
$g_s$ & 1.0 & $G_F$ & $1.16639\cdot 10^{-5}$
\end{parameters}
\end{center}

\noindent Furthermore, the values for the dependent parameters are
$\cos^2\theta_W=M_W^2/M_Z^2$ and~$\alpha=\sqrt{2}G_F\,M_W^2\sin^2\theta_W/\pi$.
For the phase space point of Tab.~\ref{tab:kin:wggg} we obtain
the numbers below.

\begin{center}\small
\begin{tabular}{|lrr|}
\hline
\multicolumn{3}{|c|}{result $u\bar{d} \to W^+ ggg$}\\
\hline
&\multicolumn{1}{l}{\GOLEM}
&\multicolumn{1}{l|}{Ref.~\cite{vanHameren:2009dr}}\\
\hline
$a_0\cdot10^7$ & $8.552735739069321$ &   \\
$c_0/a_0$    & $-36.45372625230239$ & $ -36.4536949986367$\\
$c_{-1}/a_0$ & $-34.70010131004584$ & $ -34.70007155977844$ \\
$c_{-2}/a_0$ & $-11.66666666666747$ & $-11.666656664302845$\\
\hline
\end{tabular}
\end{center}
On an Intel Core 2 i5 Laptop at 2.0\,GHz the evaluation of a single phase space
point took about 2.5\,s for $u\bar{d} \to e^+\nu_e ggg$ and about 7.5\,s for 
on-shell W's without decay.
The code was compiled with \texttt{gfortran -02}.

%% file: wBB.tex
The process $u {\bar d} \to W^+ b\overline{b}$, with an on-shell $W$-boson, has been studied in 
\cite{FebresCordero:2006sj}, while the effects of the $W$-decay have been
recently accounted for in
\cite{Badger:2010mg}, and implemented within MCFM.
We consider the latter process, and compare the renormalised amplitude evaluated by MCFM. 
The $b$-quark is treated as massive in all diagrams
except in the vacuum-polarisation like contributions.

\begin{table*}[ptb]
\begin{center}
\begin{kinematics}
$u$ & 76.084349979114506 & 0.0 & 0.0 & 76.084349979114506 \\
$\bar{d}$ & 1998.0331337409114 & 0.0 & 0.0 & -1998.0331337409114 \\
$\nu_e$ & -953.55303294091811   &  955.01676368653477  &  50.025808060592873 &  17.060211586132972  \\
$e^+$ & -190.20402007017753  & 194.22279012023398  &   4.3588877692445251 &   39.063065018596490 \\
$b$   & -417.39085287123652  & 468.23544715890415  &   208.22173996408185 &   40.625785184424117  \\
$\bar{b}$  & -360.80087787946474  & 456.64248275435313  &  -262.60643579391922 &  -96.749061789153586
\end{kinematics}
\end{center}
\caption{Kinematic point used in $u\bar{d}\to W^+b\bar{b}$.}
\label{tab:kin:wBB}
\end{table*}

\begin{center}
\begin{parameters}
$\mu$ & $80.0$ & $g_s$ & 1 \\
$m_t$ & 172.5 & $m_b$ & 4.75 \\
$M_Z$ & 91.1876 & $M_W$ & 80.419 \\
$\Gamma_W$ & 2.1054 & $G_F$ & 0.0000116639 \\
$V_{ud}$ & 0.975 & \\
\end{parameters}
\end{center}

\noindent Using the above parameters and the kinematics
given in Tab.~\ref{tab:kin:wBB} we obtain the following results.
\begin{center}
\small
\begin{tabular}{|lrr|}
\hline
\multicolumn{3}{|c|}{result $u\bar{d} \to \nu_e e^+ b \bar{b}$}\\
\hline
&\multicolumn{1}{l}{\GOLEM}
&\multicolumn{1}{l|}{MCFM-6.0}\\
\hline
$a_0\cdot10^7$ & $1.884434667673654$ & $ 1.88443466774536441$\\
$c_{0}/a_0$  & $ 41.21712989438873$ & $ 41.217129894410029 $ \\
$c_{-1}/a_0$ & $ 26.60367070701196$ & $ $ \\
$c_{-2}/a_0$ & $ -2.666666666666624$ & $ $ \\
\hline
$IR_{-1}$ & $ 26.60367070701218$ & $ $ \\
$IR_{-2}$ & $ -2.666666666666667$ & $ $ \\
\hline
\end{tabular}
\end{center}
\medskip

The evaluation of a single phase space
point took $9.12$\,ms on a 2\,GHz processor.
The code was compiled with \texttt{gfortran -O2}.

%% file: conclusion.tex
We have presented the program package \GOLEM{} 
which produces, in a fully automated way, the code required to perform 
the evaluation of one-loop matrix elements for multi-particle processes.
The program is publicly available at
\url{http://projects.hepforge.org/gosam/} 
and can be used to calculate one-loop amplitudes 
within QCD, electroweak theory, or other models 
which can be imported  via an interface to
 LanHEP  and UFO, 
 also included in the release. 
 Monte Carlo programs for the real radiation can be easily linked 
 through the BLHA interface.
 
\GOLEM{} is extremely flexible, allowing for both
uni\-ta\-ri\-ty-\hspace{0pt}based
reduction
at integrand level and traditional tensor reduction, or even for a combination of the two 
approaches when required. 
The amplitudes are generated in terms of Feynman diagrams 
within the dimensional regularisation scheme, and optionally 
the calculation can be carried out either in the \tHV{} or in the 
dimensional reduction variant. 
The user can choose among different libraries for the master integrals, and 
the setup is such that  other libraries can be linked easily.

The calculation of the rational terms is very modular and can proceed either 
along with the same numerical reduction as the rest of the amplitude, 
or independently, before any reduction,  
by using analytic information on the integrals which can potentially give rise to 
a rational part.
In the current version of the code, UV-renormalisation counterterms are provided 
for QCD corrections only. Further improvements concerning the full automatisation of electroweak corrections are planned.

Different systems to detect and rescue numerical instabilities are implemented, 
and the user can switch between them without having to re-generate the source code.
Due to a careful organisation of the calculation both at the code generation stage 
and at the reduction stage, the runtimes for multi-particle amplitudes are very satisfactory.
Moreover, the \GOSAM{} generator can also produce codes for processes
that include intermediate states with complex masses.

 Within the context of the automated matching of Monte Carlo programs 
 to NLO virtual amplitudes, 
 \GOSAM{} can be used as a module to produce 
 differential cross sections for multi-particle processes 
 which can be compared directly to experiment. 
 Therefore we believe that \GOSAM{} can contribute to the goal of 
 using NLO tools as a standard framework for the LHC data analysis
at the TeV scale.

%% file: examples.tex
In the following we give results for the processes listed in the {\tt examples}
directory.
Unless stated otherwise, we assume that the coupling constants
($e$ and $g_s$ in the standard model)
 have been set to one in the input card.
 The conventions for the returned  numbers $(a_0, c_0, c_{-1}, c_{-2})$ are as stated in
 Section~\ref{ssec:conventions}.
Dimensionful parameters are understood to be in powers of GeV.

As an illustration of the potential of \GOSAM{}, we display in Table~\ref{timecode}  the timings required by a
wide list of benchmark processes. The first value provided in the table is the time required for the
code generation (\emph{Generation}, given in seconds): we remind the reader
that this operation only needs to be performed once per process.
The second value is the timing for the full calculation of the amplitude
at one phase-space point (\emph{Evaluation}, in milliseconds).
Results are obtained with an Intel(R) Core(TM) i7 CPU 950  @ 3.07GHz.

\begin{table}[h]
\bcen
{\small \begin{tabular}{|l|r|r|}
\hline
Process                                          &
\multicolumn{1}{|c|}{Generation [s]} &
\multicolumn{1}{|c|}{Evaluation [ms]} \\
\hline
$bg\to Hb$                                       &      236&  2.49\\
$d\bar{d}\to t\bar{t}$                           &      341&  4.71\\
$d\bar{d}\to t\bar{t}\,\text{(DRED)}$            &      324&  4.05\\
$dg\to dg$                                       &      398&  3.08\\
$dg\to dg\,\text{(DRED)}$                        &      402&  3.28\\
$e^+e^-\to t\bar{t}$                             &      221&  1.27\\
$e^+e^-\to t\bar{t}\,\text{(LanHEP)}$            &      180&  1.27\\
$e^+e^-\to u\bar{u}$                             &      122&  0.65\\
$gg\to gg$                                       &      525&  1.69\\
$gg\to gg\,\text{(DRED)}$                        &      428&  1.66\\
$gg\to gg\,\text{(LanHep)}$                      &     1022&  1.70\\
$gg\to gZ$                                       &      529&  15.18\\
$gg\to t\bar{t}$                                 &     1132&  24.65\\
$gg\to t\bar{t}\,\text{(DRED)}$                  &      957&  30.13\\
$gg\to t\bar{t}\,\text{(UFO)}$                   &     1225&  29.45\\
$H\to\gamma\gamma$                               &      140&  0.24\\
$gb\to e^-\bar{\nu}_et$                          &      337&  2.89\\
$u\bar{d}\to e^-\bar{\nu}_e$                     &       71&  0.09\\
$u\bar{d}\to e^-\bar{\nu}_eg$                    &      154&  1.15\\
$u\bar{u}\to d\bar{d}$                           &      186&  2.06\\
$\bar{u}d\to W^+W^+\bar{c}s$                     &     1295&  17.37\\
$\gamma\gamma\to\gamma\gamma$                    &      597&  6.08\\
\hline
\end{tabular} } \label{timecode}
\caption{Time required for code generation and calculation of one phase-space point. 
The results were obtained with an Intel(R) Core(TM) i7 CPU 950  @ 3.07GHz.
The time for the evaluation of a phase space point is taken as the average
of the time obtained from the evaluation of 100 random points generated using
RAMBO~\cite{Kleiss:1985gy}, where the code was compiled using \texttt{gfortran}
without any optimisation options. The generation of the $R_2$ term was
set to \texttt{explicit}.
}
\ecen
\end{table}

\subsection{How to run the examples}
The example directories only define the system independent part of the
setup. All settings which  are defined in the
file \texttt{system.rc} (see Section~\ref{sec:usage}) must be put either in a file called
\texttt{\$HOME/.gosam} or in the file \texttt{setup.in} in the
\GOLEM{} \texttt{examples/} directory. A script \texttt{runtests.sh}
is provided to generate, compile and run the test programs.
The names of the directories respectively examples to be run should be specified at the
command line, e.g.
\begin{lstlisting}
./runtests.sh eeuu bghb
\end{lstlisting}
If the script
is invoked without arguments it will loop over all subdirectories.
A second script, \texttt{summarize.sh}, can be used in order to collect
the test results and print a summary to the screen.
The command
\begin{lstlisting}
./summarize.sh
\end{lstlisting}
will produce an output like the following one.
\begin{lstlisting}
+ bghb (succeeded)
+ eeuu (succeeded)
grep: ./ddtt/...: No such file ...
\end{lstlisting}
The examples $e^+e^-\to t\bar{t}$ have an explicit dependence
on the \GOLEMVC{} library and will therefore fail if the
extension \texttt{golem95} is not added.

\subsection{\texorpdfstring{$e^+e^-\to u\overline{u}$}{%
e+ e- -{}-> u u-bar}}

The following parameters and momenta have been used to produce the numerical result:

\begin{center}
\begin{kinematics}
$e^+$& $E$ & 0&0&$E$\\
$e^-$& $E$ & 0&0&-$E$\\
$u$&$E$&$E\,\sin\theta\sin\phi$&$E\,\sin\theta\cos\phi$&$E\,\cos\theta$\\
$\bar{u}$&$E$&-$E\,\sin\theta\sin\phi$&-$E\,\sin\theta\cos\phi$&-$E\,\cos\theta$\\
\end{kinematics}

\bigskip

\begin{parameters}
$E$ & 74.7646520969852&$\mu^2$ & $4\,E^2$ \\
$\phi$&2.46&$\theta$&1.35\\
$M_Z$ & 91.1876 &$\Gamma_Z$ & 2.4952 \\
$M_W$ &$\cos\theta_w\,M_Z$&$\sin\theta_w$&0.47303762\\
\end{parameters}

\bigskip

\begin{tabular}{|lrr|}
\hline
\multicolumn{3}{|c|}{\small result $e^+e^-\to u\overline{u}$}\\
\hline
&\multicolumn{1}{l}{\small\GOLEM}
&\multicolumn{1}{l|}{\small analytic}\\
\hline
\small $a_0$ & 3.7878306213027528 &\\
\small $c_0/a_0$ & \small 1.86960440108932 $\times \,C_F$ &
\small $(\pi^2-8)\times C_F$\\
\small $c_{-1}/a_0$ & \small $-3.0000000000000 \times C_F$ &
\small $-3 \times C_F$ \\
\small $c_{-2}/a_0$ & \small $-2.0000000000000 \times C_F$ &
\small $-2 \times C_F$ \\
\hline
\end{tabular}
\end{center}

\subsection{\texorpdfstring{$e^+e^-\to t\overline{t}$}{e+ e- -{}-> t t-bar}}

This example has been produced twice: once with the default model file and once with
a model file imported from  LanHEP\,\cite{Semenov:2010qt}.
Thus it also can serve as an example of how to import
model parameters from LanHEP.
The result is given in dimensional reduction, and no renormalisation terms are included.

\begin{table*}[ptb]
\begin{center}
\begin{kinematics}
$e^+$& 74.7646520969852& 0.& 0.& 74.7646520969852\\
$e^-$& 6067.88254935176& 0.& 0.& -6067.88254935176\\
$t$& 5867.13826404309&  16.7946967430656&  169.437140279981& -5862.12966020487\\
$\bar{t}$& 275.508937405653& -16.7946967430656& -169.437140279981& -130.988237049907
\end{kinematics}
\end{center}
\caption{Kinematic point used in $e^+e^-\to t\bar{t}$}
\label{tab:kin:eett}
\end{table*}

\begin{center}
\begin{parameters}
$M_Z$ & 91.1876 &$\Gamma_Z$ & 2.4952 \\
$M_W$ &$\cos\theta_w\,M_Z$&$\sin\theta_w$&0.47303762 \\
$m_t$ & 172.5& $\mu^2$ & $m_t^2$\\
\end{parameters}
\end{center}

\noindent The following results are obtained with the above parameters
and the kinematic point of Tab.\,\ref{tab:kin:eett}.

\begin{center}\small
\begin{tabular}{|lrr|}
\hline
\multicolumn{3}{|c|}{result $e^+e^-\to t\overline{t}$}\\
\hline
&\multicolumn{1}{l}{\small\GOLEM}
&\multicolumn{1}{l|}{\small analytic}\\
\hline
$a_0$ & 6.3620691850584166&6.3620691850631061\\
$c_0/a_0$ & 13.182472828297422 & 13.182472828302023\\
$c_{-1}/a_0$ & 12.211527682024421 & 12.211527682032367\\
$c_{-2}/a_0$ & 0. & 0. \\
\hline
\end{tabular}
\ecen

\subsection{\texorpdfstring{$u\overline{u}\to d\overline{d}$}{%
u u-bar -{}-> d d-bar}}

This example has been produced twice: once in the \tHV{} (HV) scheme  and once
with dimensional reduction (DRED). Only the result in the HV scheme will be listed below,
for the DRED calculation see the directory {\tt uudd\_dred}.

\begin{table*}[ptb]
\begin{center}
\begin{kinematics}
$u$&102.6289752320661& 0& 0&  102.6289752320661\\
$\bar{u}$&102.6289752320661& 0& 0& -102.6289752320661\\
$d$&102.6289752320661& -85.98802977488269& -12.11018104528534&  54.70017191625945\\
$\bar{d}$&102.6289752320661&  85.98802977488269& 12.11018104528534& -54.70017191625945
\end{kinematics}
\end{center}
\caption{Kinematic point used in $u\bar{u}\to d\bar{d}$.}
\label{tab:kin:uudd}
\end{table*}

\begin{center}
\begin{parameters}
$\mu$ & 91.188 & $N_f$&2\\
\end{parameters}
\end{center}

\noindent Using the above parameters and the phase space point of
Tab.\,\ref{tab:kin:uudd} we obtain the following numbers.
\begin{center}\small
\begin{tabular}{|lrr|}
\hline
\multicolumn{3}{|c|}{result $u\overline{u}\to d\overline{d}$}\\
\hline
&\multicolumn{1}{l}{\GOLEM{}(HV)}
&\multicolumn{1}{l|}{Ref.~\cite{Ellis:1985er}}\\
\hline
$a_0$ &0.28535063700913421&0.28535063700913416 \\
$c_0/a_0$& -2.7940629929270155& -2.7940629929268876 \\
$c_{-1}/a_0$& -6.4881359148866604&-6.4881359148866391 \\
$c_{-2}/a_0$& -5.3333333333333&  -5.3333333333333 \\
\hline
\end{tabular}
\ecen

\subsection{\texorpdfstring{$gg\to gg$}{g g -{}-> g g}}
This example has been produced both with the default model file and with
a model file imported from  LanHEP.
Further, it has been calculated in the \tHV{} scheme  and in the
dimensional reduction scheme. Only the results in the \tHV{} scheme
are listed below, for further details please see the subdirectories
{\tt gggg\_dred} and {\tt gggg\_lhep}. The result is for the helicity configuration
$g(+)g(+)\to g(-)g(-)$, and pure Yang-Mills theory, i.e. fermion loops are not included.

\begin{table*}[ptb]
\begin{center}
\begin{kinematics}
$p_1$&220.9501779577791& 0& 0&  220.9501779577791\\
$p_2$&220.9501779577791& 0& 0&  -220.9501779577791\\
$p_3$&220.9501779577791&119.9098300357375&  183.0492135511419& -30.55485589367430\\
$p_4$&220.9501779577791& -119.9098300357375&-183.0492135511419&  30.55485589367430
\end{kinematics}
\end{center}
\caption{Kinematic point used in $gg\to gg$.}
\label{tab:kin:gggg}
\end{table*}

\begin{center}
\begin{parameters}
$\mu^2$ & 442  & $N_f$&0\\
$\alpha_s$ & 0.13 &&
\end{parameters}
\end{center}

\noindent
Evaluating the amplitude for above parameters and the
phase space point given in Tab.\,\ref{tab:kin:gggg}
we obtain the following results.

\begin{center}\small
\begin{tabular}{|lrr|}
\hline
\multicolumn{3}{|c|}{result $gg\to gg$}\\
\hline
&\multicolumn{1}{l}{\GOLEM{}(HV)}
&\multicolumn{1}{l|}{Ref.~\cite{Binoth:2006hk}}\\
\hline
$a_0$ & 14.120983050796795 & 14.120983050796804\\
$c_0/a_0$ & -124.0247557942351 & -124.02475579423495\\
$c_{-1}/a_0$ & 55.003597347101078 & 55.003597347101035 \\
$c_{-2}/a_0$ & -12.00000000000000 & -12. \\
\hline
\end{tabular}
\end{center}

\subsection{\texorpdfstring{$gg\to gZ$}{g g -{}-> g Z}}

As this process has no tree level amplitude,
the result is for the one-loop amplitude squared.

\begin{table*}[ptb]
\begin{center}
\begin{kinematics}
$g$&100&0&0&100\\
$g$&100&0&0&-100\\
$g$&79.2120540156&3.65874234516586&- 25.1245942606679&75.0327786308013\\
$Z$&120.7879459844&-3.65874234516586&25.1245942606679&- 75.0327786308013
\end{kinematics}
\end{center}
\caption{Kinematic point used in $gg\to gZ$.}
\label{tab:kin:gggz}
\end{table*}

\begin{center}
\begin{parameters}
$\mu^2$ & $s_{12}$ & $\alpha_s$&1\\
$M_Z$ & 91.1876 & $\Gamma_Z$ &  0 \\
$\sin\theta_w$&0.4808222 & $M_W$ &$\cos\theta_w\,M_Z$\\
$N_f$&2 & &
\end{parameters}
\end{center}

\noindent
With the above parameters and the kinematics given
in Tab.\,\ref{tab:kin:gggz} we obtain the following result.
\begin{center}\small
\begin{tabular}{|lrr|}
\hline
\multicolumn{3}{|c|}{result $gg\to gZ$}\\
\hline
&\multicolumn{1}{l}{\GOLEM}
&\multicolumn{1}{l|}{Ref.~\cite{vanderBij:1988ac}}\\
\hline
$a_0$ & - & -\\
$\vert\mathcal{M}\vert_{\text{1-loop}}^2$ & 0.1075742599502829 &0.10757425995048300 \\
\hline
\end{tabular}
\ecen

\subsection{\texorpdfstring{$d\overline{d}\to t\overline{t}$}{%
d d-bar -{}-> t t-bar}}

This example has been calculated in the \tHV{} scheme  and in the
dimensional reduction scheme. Only the results in the \tHV{} scheme
are listed below, for the renormalised amplitude with $N_f=5$ and the top mass
renormalised on-shell.

For further details please see the subdirectories
{\tt ddtt}  and {\tt ddtt\_dred}.

\begin{table*}[ptb]
\begin{center}
\begin{kinematics}
$d$&74.7646520969852& 0& 0& 74.7646520969852\\
$\overline{d}$&6067.88254935176& 0& 0& -6067.88254935176\\
$t$&5867.13826404309&  16.7946967430656& 169.437140279981& -5862.12966020487\\
$\overline{t}$&275.508937405653& -16.7946967430656&-169.437140279981& -130.988237049907
\end{kinematics}
\end{center}
\caption{Kinematic point used in $d\overline{d}\to t\overline{t}$.}
\label{tab:kin:ddtt}
\end{table*}

\begin{center}
\begin{parameters}
$m_t$ & 172.5&$\mu^2$ &$m_t^2$\\
$\alpha_s$&1& $N_{f}$&5\\
\end{parameters}
\end{center}

\noindent
With the above parameters and the kinematics given
in Tab.\,\ref{tab:kin:ddtt} we obtain the following results.

\begin{center}\small
\begin{tabular}{|lrr|}
\hline
\multicolumn{3}{|c|}{result $d\overline{d}\to t\overline{t}$}\\
\hline
&\multicolumn{1}{l}{\GOLEM{}(HV)}
&\multicolumn{1}{l|}{Ref.~\cite{Campbell:1999ah,Campbell:2000bg} (MCFM)}\\
\hline
$a_0$ & 0.43024349783870747 & 0.43024349783867882\\
$c_0/a_0$ & -22.526901042662193 & -22.526901042658068\\
$c_{-1}/a_0$ & 10.579577611830414 & 10.579577611830567\\
$c_{-2}/a_0$ &-2.6666666666666599  &-2.666666666666721 \\
\hline
\end{tabular}
\end{center}

\subsection{\texorpdfstring{$gg\to t\overline{t}$}{g g -{}-> t t-bar}}
The result is for the renormalised amplitude in the HV scheme.

\begin{table*}[ptb]
\begin{center}
\begin{kinematics}
$g$&137.84795086008967&   0.&  0.&   137.84795086008967\\
$g$&3161.1731634194916&   0.&  0.&  -3161.1731634194916\\
$t$&3058.6441209877348&   16.445287185144903& 165.91204201912493&  -3049.2945357402382\\
$\overline{t}$&240.37699329184659&  -16.445287185144903& -165.91204201912493&   25.969323180836145
\end{kinematics}
\end{center}
\caption{Kinematic point used in $gg\to t\overline{t}$}
\label{tab:kin:ggtt}
\end{table*}

\begin{center}
\begin{parameters}
$m_t$ & 171.2&$\Gamma_t$ &0\\
$N_f$&5&$\mu$ &71.2 \\
\end{parameters}
\end{center}

\noindent
With the above parameters and the kinematics given
in Tab.\,\ref{tab:kin:ggtt} we obtain the following results.
\begin{center}\small
\begin{tabular}{|lrr|}
\hline
\multicolumn{3}{|c|}{result $gg\to t\overline{t}$}\\
\hline
&\multicolumn{1}{l}{\GOLEM{}(HV)}
&\multicolumn{1}{l|}{Ref.~\cite{Campbell:1999ah,Campbell:2000bg} (MCFM)}\\
\hline
$a_0$ & 4.5576116986983433 & 4.5576116986983424\\
$c_0/a_0$ & 15.352143751168184 & 15.352143750919995\\
$c_{-1}/a_0$ & -27.235240992743407 & -27.235240936279297 \\
$c_{-2}/a_0$ & -6.0 & -6.0 \\
\hline
\end{tabular}
\end{center}

\subsection{\texorpdfstring{$bg\to Hb$}{b g -{}-> H b}}

For this process the mass of the b-quark is set to zero.
However, in order to have a coupling between the b-quark and the Higgs boson,
the following Yukawa coupling is implemented in the model file:
$${\cal L}_{\rm{yuk}}=Y_{Hb}\,\bar{\psi}_L\psi_R\,\phi\, , \,
 Y_{Hb}= \frac{\bar{m}_b(\mu)}{v}\;.$$

\begin{table*}[ptb]
\begin{center}
\begin{kinematics}
$b$&250&  0& 0&  250\\
$g$&250&  0& 0& -250\\
$H$&264.4& -83.84841332241601& -86.85350630148753&  -202.3197272300720\\
$b$&235.6&  83.84841332241601&  86.85350630148753& 202.3197272300720
\end{kinematics}
\end{center}
\caption{Kinematic point used in $bg\to Hb$.}
\label{tab:kin:bghb}
\end{table*}

\begin{center}
\begin{parameters}
$m_b$&0&$\bar{m}_b(\mu)$&2.937956\\
$m_H$ &120 &$v$&246.2185\\
$\mu$ &$91.188$ & &\\
\end{parameters}
\end{center}

\noindent
With the above parameters and the kinematics given
in Tab.\,\ref{tab:kin:bghb} we obtain the following results.
\begin{center}\small
\begin{tabular}{|lrr|}
\hline
\multicolumn{3}{|c|}{result $bg\to Hb$}\\
\hline
&\multicolumn{1}{l}{\GOLEM{}(HV)}
&\multicolumn{1}{l|}{Refs.~\cite{Campbell:2002zm,Hirschi:2011pa}}\\
\hline
$a_0\cdot10^7$ & 2.09926265849001642& 2.09926265848997195\\
$c_0/a_0$ & -24.131948141318752 & -24.131948141995107\\
$c_{-1}/a_0$ &11.957924609547224  & 11.957924605423791\\
$c_{-2}/a_0$ & -5.6666666666666643 & -5.6666666666666670 \\
\hline
\end{tabular}
\end{center}

\subsection{\texorpdfstring{$H\to\gamma\gamma$}{%
H -{}-> photon photon}}
The decay width $\Gamma_{H\to\gamma\gamma}$ of this loop induced
process is known analytically at lowest order.
For comparison we used the equations including the top loop and the
bosonic contribution given
in~\cite{Aglietti:2004nj,Harlander:2005rq}. The decay width can be expressed
as
\begin{equation}
\Gamma_{H\to\gamma\gamma}=\frac{G_F\alpha^2m_H^3}{128\sqrt2\pi^3}
\cdot\hat{\Gamma}(\tau_W, \tau_t)
\end{equation}
where $\tau_i=m_H^2/(4m_i^2)$ for $i=W,t$.

\begin{center}
\begin{parameters}
$m_H$&124.5&$m_t$&172.5\\
$m_W$&80.398&&
\end{parameters}

\bigskip

\small
\begin{tabular}{|lrr|}
\hline
\multicolumn{3}{|c|}{result $H\to \gamma\gamma$}\\
\hline
&\multicolumn{1}{l}{\GOLEM{}}
&\multicolumn{1}{l|}{Refs.~\cite{Aglietti:2004nj,Harlander:2005rq}}\\
\hline
$\hat{\Gamma}(\tau_W,\tau_t)$ & 3.366785118586698& 3.36678512043889\\
\hline
\end{tabular}
\end{center}

\subsection{\texorpdfstring{$u\overline{d}\to e^-\,\overline{\nu}_e$}{%
u d-bar -{}-> e- nu\_e}}
This example has been calculated in the \tHV{} scheme  and in the
dimensional reduction scheme. Only the results in the \tHV{} scheme
are listed below, for the renormalised amplitude.
In addition to a calculation with the default model file, calculations using
LanHEP\,\cite{Semenov:2010qt}  and UFO\,\cite{Degrande:2011ua}
are also contained in the examples directory.


\begin{table*}[ptb]
\begin{center}
\begin{kinematics}
$u$&100&0&0&100\\
$\overline{d}$&100&0&0&-100\\
$e^-$&100& 75.541566535633046&  30.240603423558878&   -58.128974100026611\\
$\overline{\nu}_e$&100& -75.541566535633046&  -30.240603423558878&   58.128974100026611
\end{kinematics}
\end{center}
\caption{Kinematic point used in $u\overline{d}\to e^-\,\overline{\nu}_e$.}
\label{tab:kin:udene}
\end{table*}

\begin{center}
\begin{parameters}
$\sqrt{s}$& 200 &$\mu$ & $91.1876$ \\
\end{parameters}
\end{center}

\noindent
With the above parameters and the kinematics given
in Tab.\,\ref{tab:kin:udene} we obtain the following results.
\begin{center}
\small
\begin{tabular}{|lrr|}
\hline
\multicolumn{3}{|c|}{result $u\overline{d}\to e^-\,\overline{\nu}_e$}\\
\hline
&\multicolumn{1}{l}{\GOLEM{}(HV)}
&\multicolumn{1}{l|}{Ref.~\cite{Hirschi:2011pa}}\\
\hline
$a_0$ & 1.4138127601912656 & 1.4138127601912673\\
$c_0/a_0$ & 5.4861229357937624 & 5.4861229357937660\\
$c_{-1}/a_0$ & 0.18879169932851950 & 0.18879169932852413\\
$c_{-2}/a_0$ & -2.666666666666667 &  -2.6666666666666665\\
\hline
\end{tabular}
\end{center}

\subsection{\texorpdfstring{$u\overline{d}\to e^-\,\overline{\nu}_e\,g$}{%
u d-bar -{}-> e- nu\_e g}}

We list the renormalised amplitude in the HV scheme.

\begin{table*}[ptb]
\begin{center}
\begin{kinematics}
$u$&500& 0&0& 500\\
$\overline{d}$&500& 0&0& -500\\
$e^-$&483.244841094218&   -86.3112218694181&  147.629518147233&   -451.975082051212\\
$\overline{\nu}_e$&279.253370247231&    6.62401666401929&  -5.58083951102529&    279.119009435087\\
$g$&	237.501788658551&    79.6872052053988&  -142.048678636208&    172.856072616124
\end{kinematics}
\end{center}
\caption{Kinematic point used in $u\overline{d}\to e^-\,\overline{\nu}_e\,g$.}
\label{tab:kin:udeneg}
\end{table*}

\begin{center}
\begin{parameters}
$M_W$&80.398 & $\Gamma_W$& 2.1054 \\
$\sin\theta_w$&0.4808222 &$M_Z$& $M_W/\cos\theta_w$\\
$N_f$&5 & $V_{ud}$&0.97419\\
$\mu^2$ & $s_{12}$ &&\\
\end{parameters}
\end{center}

\noindent
With the above parameters and the kinematics given
in Tab.\,\ref{tab:kin:udeneg} we obtain the following results.
\begin{center}\small
\begin{tabular}{|lrr|}
\hline
\multicolumn{3}{|c|}{result $u\overline{d}\to e^-\,\overline{\nu}_e\,g$}\\
\hline
&\multicolumn{1}{l}{\GOLEM{}(HV)}
&\multicolumn{1}{l|}{Ref.~\cite{Hirschi:2011pa}}\\
\hline
$a_0\cdot10^7$ & 2.8398509625435832 &2.8398509625435922\\
$c_0/a_0$ & -8.6052919370147745 & -8.6052919368774248\\
$c_{-1}/a_0$ & -18.722010655600936 & -18.722010655557121\\
$c_{-2}/a_0$ & -5.6666666666666 & -5.66666666666667 \\
\hline
\end{tabular}
\end{center}

\subsection{\texorpdfstring{$g\,b\to e^-\,\overline{\nu}_e\,t$}{%
g b -{}-> e- nu\_e t}}

We list the renormalised result in the dimensional reduction scheme.
\begin{table*}[ptb]
\begin{center}
\begin{kinematics}
$g$ & 1187.7086110647201&0 &0 & 1187.7086110647201\\
$b$ &2897.148136260289 & & &-2897.148136260289 \\
$e^-$ &2293.0435558834492 & 629.81047833131981&258.58120146220904 & -2189.6399870328105\\
$\overline{\nu}_e$&509.48956356743611 & 144.72113807954338&19.883362437475 & -488.098411670514\\
$t$ & 1282.3236278741238&-774.53161641086319 & -278.46456389968404&968.29887350775562 \\
\end{kinematics}
\end{center}
\caption{Kinematic point used in $g\,b\to e^-\,\overline{\nu}_e\,t$.}
\label{tab:kin:tne}
\end{table*}

\begin{center}
\begin{parameters}
$M_W$ & 80.4190 & $\Gamma_W$ & 2.04760  \\
$M_Z$ & 91.1876 & $\Gamma_Z$ & 2.49520  \\
$m_t$ & 171.2   & $\Gamma_t$ & 0  \\
$\mu$ & 71.2   & $e$ & 0.30794906326863203 \\
\end{parameters}
\end{center}

\noindent
With the above parameters and the kinematics given
in Tab.\,\ref{tab:kin:tne} we obtain the following results.
\begin{center}\small
\begin{tabular}{|lrr|}
\hline
\multicolumn{3}{|c|}{result $g\,b\to e^-\,\overline{\nu}_e\,t$}\\
\hline
&\multicolumn{1}{l}{\GOLEM}
&\multicolumn{1}{l|}{Ref.~\cite{Campbell:1999ah,Campbell:2000bg} (MCFM)}\\
\hline
$a_0\cdot10^2$ & 8.52301540675800134&  8.52301540708130106\\
$c_0/a_0$ &-79.879718568538991   & -79.879718569273024  \\
$c_{-1}/a_0$ & 26.570185488790770 & 26.570185487963364 \\
$c_{-2}/a_0$ &  -4.3333333333333401 & -4.3333333331689596\\
\hline
\end{tabular}
\end{center}

\subsection{\texorpdfstring{$\overline{u}\,d\to W^+W^+\,\overline{c}\,s$}{%
p p -{}-> W+ W+ j j}}

Results are given for the unrenormalised amplitude in the dimensional reduction scheme.

\begin{table*}
\begin{center}
\begin{kinematics}
$\overline{u}$&500& 0& 0& 500\\
$d$&500& 0& 0& -500\\
$\overline{c}$&54.2314070117999&  -7.92796656791140&  43.6912823611163& -31.1330162081798\\
$s$&214.488870161418& -98.5198083786150&  188.592247959949& -27.0607980217775\\
$e^+$&85.5312248384887&  36.1637837682033& -77.0725048002414& -8.22193223977868\\	
$\nu_e$&181.428811610043&  -171.863734086635& -5.61185898481311& -57.8599829481937\\	
$\mu^+$&82.8493010774356&  -49.8952157196287&  5.51413360058664& -65.9095476235891\\	
$\nu_\mu$&381.470385300815& 292.042940984587& -155.113300136598&  190.185277041519	
\end{kinematics}
\end{center}
\caption{Kinematic point used in $\overline{u}\,d\to W^+W^+\,\overline{c}\,s$.}
\label{tab:kin:wwcs}
\end{table*}

\begin{center}
\begin{parameters}
$\mu$ &$80$ & $N_f$&5\\
\end{parameters}
\end{center}

\noindent
With the above parameters and the kinematics given
in Tab.\,\ref{tab:kin:wwcs} we obtain the following results.
\begin{center}\small
\begin{tabular}{|lrr|}
\hline
\multicolumn{3}{|c|}{result $\overline{u}\,d\to W^+W^+\,\overline{c}\,s$}\\
\hline
&\multicolumn{1}{l}{\GOLEM}
&\multicolumn{1}{l|}{Ref.~\cite[v3]{Melia:2010bm}}\\
\hline
$a_0$ &  & \\
$c_0/a_0$ & 23.3596455167118 & 23.35965\\
$c_{-1}/a_0$ & 13.6255429251954 & 13.62554\\
$c_{-2}/a_0$ & -5.333333333333 & -5.33333 \\
\hline
\end{tabular}
\end{center}

%% file: app-r2.tex
In this Appendix we list all integrals which give rise to $R_2$ terms
as we use these expressions in their explicit construction.
We use the definition
\begin{align}
I_N^{n,\alpha;\mu_1\ldots\mu_r}(S)&=
\int\frac{\mu^{2\varepsilon}\mathrm{d}^nq}%
{i\pi^{n/2}}\frac{\hat{q}^{\mu_1}\cdots\hat{q}^{\mu_r}(\mu^2)^\alpha}{%
D_1\cdots D_ N}\\
\intertext{with}
D_l&=(q+r_l)^2-m_l^2\nonumber\\
\intertext{and}
S_{ij}&=(r_i-r_j)^2-m_i^2-m_j^2.
\end{align}

The integrals  up to $\mathcal{O}(\varepsilon)$ are
\begin{align}
\varepsilon\cdot I_1^{n,0}(S)&=-\frac12S_{11}\\
\varepsilon\cdot I_1^{n,0;\mu_1}(S)&=\frac12 S_{11} \cdot r_1^{\mu_1}\\
I_2^{n,1}(S)&=-\frac16\left(S_{11}+S_{12}+S_{22}\right)\\
\varepsilon\cdot I_2^{n,0}(S)&=1\\
\varepsilon\cdot I_2^{n,0;\mu_1}(S)&=
   -\frac12\left(r_1^{\mu_1}+r_2^{\mu_1}\right)\\
\varepsilon\cdot I_2^{n,0;\mu_1\mu_2}(S)&=
\frac16\left(%
2r_1^{\mu_1}r_1^{\mu_2}
+r_1^{\mu_1}r_2^{\mu_2}
+r_2^{\mu_1}r_1^{\mu_2}
+2r_2^{\mu_1}r_2^{\mu_2}\right)
\nonumber\\
&-\frac1{12}\hat{g}^{\mu_1\mu_2}\left(S_{11}+S_{12}+S_{22}\right)
\end{align}
\begin{align}
I_3^{n,1}(S)&=\frac12\\
I_3^{n,1;\mu_1}(S)&=-\frac16\left(r_1^{\mu_1}+r_2^{\mu_1}+r_3^{\mu_1}\right)\\
\varepsilon\cdot I_3^{n,0;\mu_1\mu_2}(S)&=\frac14\hat{g}^{\mu_1\mu_2}\\
\varepsilon\cdot I_3^{n,0;\mu_1\mu_2\mu_3}(S)&=-\frac{1}{12}
\sum_{l=1}^3\left[\hat{g}^{\bullet\bullet}r_{l}^{\bullet}\right]^{\mu_1\mu_2\mu_3}%
\\
I_4^{n,1;\mu_1\mu_2}(S)&=\frac{1}{12}\hat{g}^{\mu_1\mu_2}\\
I_4^{n,2}(S)&=-\frac16\\
\varepsilon\cdot I_4^{n,0;\mu_1\mu_2\mu_3\mu_4}(S)&=
   \frac1{4!}\left[\hat{g}^{\bullet\bullet}%
\hat{g}^{\bullet\bullet}\right]^{\mu_1\mu_2\mu_3\mu_4}\;.
\end{align}